\documentclass{article}%
\usepackage{amsmath}
\usepackage{amsfonts}
\usepackage{amssymb}
\usepackage{graphicx}%
\providecommand{\U}[1]{\protect\rule{.1in}{.1in}}
\newtheorem{theorem}{Theorem}

\newtheorem{definition}[theorem]{Definition}

\begin{document}

\title{The role of positivity and causality in interactions involving higher spin\\{\small Nuclear Physics B 941 (2019) 91-144} }
\author{{\large Bert Schroer}\\{\small CBPF, Rua Dr. Xavier Sigaud 150, 22290-180 Rio de Janeiro, Brazil }\\{\small permanent address: Institut f\"{u}r Theoretische Physik, FU-Berlin, }\\{\small Arnimallee 14, 14195 Berlin, Germany }\\{\small email: schroer@zedat.fu-berlin.de}}
\date{June 15$^{th}$ 2018}
\maketitle
\tableofcontents

\begin{abstract}
It is shown that the recently introduced positivity and causality preserving
string-local quantum field theory (SLFT) resolves most No-Go situations in
higher spin problems. This includes in particular the Velo-Zwanziger causality
problem which turns out to be related in an interesting way to the solution of
zero mass Weinberg-Witten issue. In contrast to the indefinite metric and
ghosts of gauge theory, SLFT uses only positivity-respecting physical degrees
of freedom. The result is a fully Lorentz-covariant and causal string field
theory in which light- or space-like linear strings transform covariant under
Lorentz transformation.

The cooperation of causality and quantum positivity in the presence of
interacting $s\geq1$ particles leads to remarkable conceptual changes. It
turns out that the presence of $H$-selfinteractions in the Higgs model is not
the result of SSB on a postulated Mexican hat potential, but solely the
consequence of the implementation of positivity and causality. These
principles (and not the imposed gauge symmetry) account also for the
Lie-algebra structure of the leading contributions of selfinteracting vector mesons.

Second order consistency of selfinteracting vector mesons in SLFT requires the
presence of $H$-particles; this, and not SSB, is the raison d'\^{e}tre for
$H.$

The basic conceptual and calculational tool of SLFT is the S-matrix. Its
string-independence is a powerful restriction which determines the form of
interaction densities in terms of the model-defining particle content and
plays a fundamental role in the construction of pl observables and sl
interpolating fields.

\end{abstract}

\hfill\textsl{Dedicated to Klaus Fredenhagen on the occasion of his 70th
birthday}

\section{Introduction and history of the problem}

The positivity property of quantum states guaranties the probabilistic
interpretation of quantum theory. It enters the mathematical formalism through
the identification of states with unit rays in a Hilbert space on which the
quantum observables act as operators. In quantum field theory (QFT), or more
generally for models with infinitely many degrees of freedom, it is often more
appropriate to identify states with positive linear functionals on operator
algebras. Thanks to the existence of a canonical construction\footnote{The
"reconstruction theorem" in \cite{St-Wi} is a special case of the more general
Gelfand-Naimark-Segal ("GNS") reconstruction theorem \cite{Haag}.} this
formulation in terms of expectation values permits a return to the more common
Hilbert space setting.

Its validity in quantum mechanics is guarantied by Heisenberg's canonical
quantization of positions and momenta in conjunction with the von Neumann
uniqueness theorem which insures that irreducible representations of the
Heisenberg commutation relations are unitarily equivalent to the
Schr\"{o}dinger representation. Born's identification of the absolute square
of the Schr\"{o}dinger wave function with the probability density for finding
a particle at a particular position connects positivity with spatial localization.

This situation undergoes significant changes in relativistic QFT where the
positivity of field-quantization looses its "von Neumann protection" in the
presence of higher spin $s\geq1$. The first such clash with positivity was
noticed by Gupta and Bleuler who observed that quantized \textit{massless}
vector potentials are incompatible with Hilbert space positivity. In the
absence of interactions it is straightforward to restore positivity by passing
from potentials to field strengths, but the use of local gauge invariance to
preserve at least part of positivity in the presence of interactions leads to
a loss of important physical operators and states.

This includes in particular all interacting fields which interpolate
charge-carrying particles in the sense of large time scattering theory. Such
\textit{interpolating fields} play an indispensable role in connecting the
causal localization- and quantum positivity- principles of QFT with observed
scattering properties of particles. Their absence in quantum gauge theory (GT)
is accompanied by a loss of mathematical tools of functional analysis. The
proofs of structural properties as TCP and Spin\&Statistics theorems use
Hilbert space positivity in an essential way and have no substitute in
indefinite metric Krein spaces. This reduces the use of GT to perturbative
rules for dealing with indefinite metric- and ghost- degrees of freedom (the
BRST formalism)

Positivity-obeying massive tensor potentials and their spinorial counterpart
are provided by Wigner's unitary representation theory of positive energy
particle representations of the (covering of the) Poincar\'{e} group, but they
come with an increase of their short distance scale dimension\footnote{It is
most conveniently obtained from property of the field's 2-ptfct $x\rightarrow
\lambda x$ for $\lambda\rightarrow0.$} with spin $d_{sd}=s+1~$which prevents
their use in renormalized perturbation theory involving fields with higher
spin $s\geq1.$ It turns out that this worsening of short distance behavior
with increasing spin is accompanied by a $m^{-s}$ divergence for small masses.
Hence a formulation of QED in terms of positivity-maintaining point-local
potentials is not possible.

In his well-known monograph Weinberg presents a systematic construction of the
intertwiner functions which relate Wigner's spin $s$ momentum space particle
creation and annihilation operators $a^{\#}(p,s)$ associated with the unitary
($m,s$) representations with covariant pl free fields which act in the
Wigner-Fock Hilbert space of the Wigner operators \cite{Wein}. This
interesting section in his book remained a torso since the (with increasing
$s$) worsening short distance scale dimension of point-local fields prevents
their use in renormalized perturbation theory as soon as $s\geq1$.

In the main part of his book Weinberg uses the positivity-violating (but
renormalizability-improving) gauge theoretic setting as obtained by Lagrangian
quantization in which a perturbative inductive argument secures the positivity
of gauge invariant operators. For this reason one does not find GT in
presentations of nonperturbative QFT.

The independence of short distance dimensions of quantized gauge fields from
spin/helicity is a consequence of the spin independence of the classical
dimension $d_{cl}=4$ of Lagrangians. For $s=0,1/2$ these fields agree with
those obtained from the Wigner-Weinberg construction, but for $s\geq1$ the
equality of the short distance dimension with the classical dimension in terms
of mass units ("engineering" dimension), namely $d_{sd}=d_{cl}=1$ for integer
and $3/2$ in case of half-integer spin, comes with an improvement of
renormalizability at the price of the presence of unphysical degrees of freedom.

Whereas in older work \cite{Fro} positivity problems for propagators of higher
spin fields in GT have been at least partially addressed, more recent
publications (\cite{Vas} \cite{Bek} and papers cited therein) are mainly
concerned with classical geometric aspects of the Lagrangian gauge formalism
for which these problems can be ignored.

The setting of string-local quantum field theory (SLFT) in the present article
overcomes this conceptual gap between GT and costructions of fields based on
Wigner's representation theory by providing a positivity maintaining causal
perturbative QFT formalism which includes the important physical interpolating
fields of particles whose large-time properties account for a unitary S-matrix
and which are missing in GT. After almost 70 years of GT this amounts to a
paradigmatic shift which does not only affect renormalized perturbation theory
but also requires to extend the nonperturbative setting of "axiomatic QFT" as
presented in \cite{St-Wi}.

A convenient starting point is to recall the construction of positivity
obeying quantum fields $\Psi_{a}~$in Weinberg's intertwiner formulation (for
simplicity for massive tensor potentials):%
\begin{align}
\Psi_{\alpha}(x)  &  =\int(\sum_{s_{3}=-s}^{s}e^{ipx}v_{\alpha,s_{3}%
}(p)a^{\ast}(p,s_{3})+\hbox{h.c.})d\mu_{m}(p)+h.c.,~~\label{W}\\
&  with~\ d\mu_{m}(p)=\theta(p_{0})\delta(p^{2}-m^{2})d^{4}p\nonumber
\end{align}
The intertwiner functions $v(p)$ convert the Wigner creation/annihilation
operators $a^{\#}(p,s_{3})$ into covariant fields $\Psi_{\alpha}$; their
calculation uses only group theory \cite{Wein}. They come with two indices,
the $s_{3}$ which runs over the $2s+1$ values of the third component of the
physical spin, and a tensor index $\alpha=(\mu_{1},..\mu_{s})$ which refers to
the $4s$ dimensional tensor representation of the Lorentz group of tensor
degree $s$. The extension to fermions is straightforward but not needed for
the problems addressed in the present work.

The momentum space Wigner creation/annihilation operators $a^{\#}(p,s_{3}),$
and hence also the covariant fields $\Psi_{\alpha}.$ act in the Wigner-Fock
Hilbert space obtained from the 1-particle Wigner representation by "second
quantization"\footnote{Note the difference to the standard use of
"quantization" (in the words of Ed Nelson: "second quantization is a functor,
wheres quantization is an art").} . Looking at the explicit form of the
intertwiners and calculating the two-point function (2-ptfct) of $\Psi$ one
finds that the latter scales as $\lambda^{-2(s+1)}$ for $x\rightarrow\lambda
x$ in the limit of small distances $\lambda\rightarrow0$ which leads to
assigning the short distance dimension $d_{sd}=s+1$ to the point-local (pl)
$\Psi.$

This construction via intertwiners permits much more flexibility than
Lagrangian or functional integral quantization in converting the unique
$(m,s)$ Wigner operators into fields with different prescribed covariance- and
causal localization- properties than quantization; includes in particular
string-local (sl) covariant quantum fields with improved ($0<d_{sd}<s+1$)
short distance properties. such fields are localized on causally separable
(i.e. allowing relative causal positioning)~semiinfinite space- or light-like
strings (rays) $\mathcal{S}=x+\mathbb{R}_{+}e,$ $e^{2}=-1$ or~$0.$ Whereas in
Weinberg's construction the covariance under Lorentz transformations is
sufficient since causality , the construction of sl fields requires the
\textit{direct use of causal localization}. The resulting covariant fields
extend the linear part of the pl relative causality class to sl (with pl
considered as a special case of sl) and Wick-ordered products thereof
constitute the nonlinear members.

This huge set of sl free quantum fields associated to an irreducible Wigner
representation contains in particular a \textit{sl tensor fields which is
linearly associated with its pl counterpart}. This sl tensor field appears
together with $s$ \textit{escort fields} with lower tensor degrees
\cite{E-M}\cite{Bey}$.$ Escorts are reminiscent of negative metric
St\"{u}ckelberg fields in gauge theory, except that they do not add unphysical
degrees of freedom to the physical $a^{\#}(p,s_{3})$ Wigner operators but only
differ in their intertwiner functions.

Positivity and hence the unitarity of the S-matrix in the resulting
string-local QFT (SLFT) is automatic\footnote{The causal separation properties
of sl fields are more than enough for deriving linked cluster fall-off
properties and insure the $e$-independence of the (on-shell) S-matrix.} (no
Nobel-prize worthy hard work as in gauge theory) and the chances to solve
age-old infrared problems (large time scattering theory in QED, QCD
confinement,..) are significantly enhanced \footnote{Recently Rehren showed
that infinite spin fields can be obtained in terms of appropriatly defined
Pauli-Lubanski limits of finite spin escort fields \cite{PL}.}. One
prerequisite is the substitution of nonexistent positivity-maintaining pl
potentials by sl counterparts and the according to the Weinberg-Witten No-Go
theorem \cite{W-W} missing $h\geq1~$sl current and stress-energy tensors in
\cite{MRS2} by suitably defined conserved sl substitutes. The smooth passing
from massive sl two-point functions with $2s+1$ degrees of freedom to their
massless two-component helicity counterpart leads to a profound (indefinite
metric- and ghost-free) understanding of the D-V-Z discontinuity problem
\cite{DVZ,Z}.

An important step in the development of$~$(SLFT) was the\textit{ construction
of fields for the class of massless infinite spin Wigner representations}%
$~$for which Yngvason's 1970 No-Go theorem excluded pl fields \cite{Y}. For
this class Weinberg's group theoretic method is without avail; one rather had
to resort to ideas from modular localization \cite{MSY1,MSY2,M}. This paved
the way for the construction of the simpler finite spin sl free fields,
including the use of their short distance lowering and hence renormalization
improving properties in interactions.

In the same work it was also realized that \textit{finite} spin/helicity sl
fields can also be obtained in a more direct way by integrating pl fields
along semi-infinite lines. This direct construction is particularly useful for
those sl potentials and their escorts which are linearly related to the pl
spin $s\geq1$ potentials. Different from the pl potentials which diverge in
the massless limit, the corresponding sl potentials pass to corresponding
finite helicity potentials which have no pl counterpart.

An important support for a string-extended QFT comes from work by Buchholz and
Fredenhagen who used the setting of algebraic QFT \cite{Bu-Fr} to show that
models with particle states separated by spectral gaps which fulfill certain
consistency properties with respect to the local observables always contain
interpolating operators localized in arbitrarily narrow spacelike cones (whose
cores are strings). Perturbative SLFT is more specific by showing that in the
presence of $s\geq1$ particles positivity together with causal localizability
leads to noncompact causal localization whose tightest localized covariant
generating fields are string-local.

The combinatorial nature of perturbation theory per se does not require
positivity and works also for gauge theory, but without positivity provided by
a Wigner-Fock Hilbert space the quantum theory remains incomplete. SLFT
reveals among other things that several limitations of gauge theory which are
the cause of certain No-Go theorems of which the best known is the
aforementioned Weinberg-Witten No-Go theorem (for a recent survey see
\cite{Ra-Ta}) are converted into Yes-Go statements in SLFT \cite{MRS2}.

SLFT is the only formulation in which state-creating interpolating fields are
separated from observables by spacetime localization properties. Whereas in
the absence of interactions the localization of free fields associated to a
Wigner representation ("kinematic localization") may be chosen at will, that
of \textit{interacting} fields in SLFT is determined by the particle content
of the interacting theory: observables are pl and interpolating fields are sl.

The space- or light-like interpolating fields can be placed in spacelike
separated positions which is a prerequisite for the application of the LSZ
scattering theory. The particle states in which the expectation values of
observables are measured are constructed in terms of suitably defined large
time asymptotic limits to the vacuum. A theorem of large-time scattering
theory insures that the dependence on the interpolating operator in the large
time limit is contained in its vacuum-to-one particle matrixelement which is
then removed by passing to the correctly normalized particle states. This
implies in particular that the $e$-dependence of sl fields \textit{does not
affect particles and their scattering matrix}.

The so-called cluster decomposition property of correlation functions of
fields plays an important role in the derivation of scattering properties. It
is a consequence of a mass gap and the existence of an arbitrary large number
of sl fields in relative spacelike position. This applies to spacelike strings
and (with some stretch of geometric imagination) also holds for lightlike
strings; it is however violated for timelike strings. In the present work the
terminology "nonlocal" is avoided since its historical connotation may hinder
to see that these fields can be brought into causally separated positions.

The reader is also reminded that the terminology "interaction density" instead
of "interacting part of a Lagrangian" is not nitpicking; apart from
interactions between $s<1$ particles, interaction densities
\textit{constructed from sl Wigner fields} are never interacting parts of Lagrangian.

Another important point which requires attention is the fact that the
kinematical sl localizations of free fields in terms of line integrals over
$s\geq1$ pl fields only serves to \textit{construct interaction densities
whose S-matrix is independent on string directions}. As mentioned the physical
localization of the corresponding interacting fields is dynamical and
generally different from that of their free counterpart used in the
construction of the interaction density and the S-matrix.

A surprising property of SLFT is that in the presence of $s\geq1$ particles
its positivity and localization properties determine a unique model in terms
of its particle content whenever such a theory exists. This is a result of the
strong restriction which the string-independence of the S-matrix exerts on
interaction densities.

The quest for an \textit{intrinsic} formulation in which the umbilical chord
to classical field theory provided by quantization has been cut is almost as
old as QFT. In the first (still pre-renormalization) presentation of quantum
electrodynamics at an international conference in 1929 \cite{Charkov} Pascual
Jordan expressed this in the form of a plea for an intrinsic understanding of
QFT which avoids the use of "(quasi)classical crutches"; a decade later his
former collaborator Eugen Wigner took the first step in his famous
classification of relativistic particles \cite{Wig}.

The second step was taken two decades later by Rudolf Haag \cite{Haag} when he
proposed an intrinsic formulation of QFT in terms of "causal nets of algebras"
in which Wightman fields at best play the role of "coordinatizations" (in
analogy to the use of coordinates in geometry).

With the arrival of the covariant formulation of quantized electrodynamics in
the 50s, Jordan's dictum and its partial realization in Wigner's
classification of noninteracting particles faded into the background; the new
covariant computational rules of quantized electrodynamics took a firm hold
and as a result the first covariant QFT was a positivity-violating gauge theory.

A somewhat unexpected aspect of these first successful calculations was the
contrast between the precision of the experimentally verified perturbative
results and the robustness of the calculated results against the use of quite
different cutoff- and regularization- prescriptions, or even against different
ways of implementing Lagrangian quantization (Gell-Mann--Low, Feynman path
integrals, Bogoliubov's generating S-functional).

The presence of gauge theoretic indefinite metric degrees of freedom in
interaction densities involving $s\geq1$ particles led to conceptual problems.
A formal proposal to overcome these shortcomings was made by Jordan \cite{Jo}.
It consisted in replacing the gauge dependent matter field\footnote{Jordan
used these fields for a pure algebraic derivation of Dirac's geometric
magnetic monopole quantization \cite{mo}.} by the formally gauge invariant
string-local composite field
\begin{equation}
\Psi(x)=\psi^{K}(x)\exp ig\int_{0}^{\infty}A_{\mu}^{K}(x_{0},\dots
,x_{3}+\lambda)d\lambda
\end{equation}
~where the $K~$refers to the gauge dependent Lagrangian field which acts in an
indefinite metric Krein space which in addition to physical degrees of freedom
contains also indefinite metric quanta (scalar and longitudinal photons).

After the discovery of renormalized perturbation theory Mandelstam used such
representation as a starting point in his attempt to construct a perturbation
theory which avoids the use of potentials in favor of working directly with
gauge invariant fields \cite{Man}. Subsequently Steinmann \cite{Stein} studied
the problem of recovering positivity by constructing such fields $\Psi~$in
higher order perturbation theory. Different proposals to recover positivity
can be found in \cite{Mor-Stro}. The constructions of such formally gauge
invariant composite fields and their renormalization requires a lot of
additional work and is of little interest unless it leads to new physical insights.

The SLFT perturbation theory in the present paper uses sl potentials with the
$s$-independent short distance dimension $d_{sd}=1$ which "live" in a physical
Wigner-Fock particle space$.~$The starting point is the observation that there
exist sl vector potentials $A_{\mu}(x,e)$ localized on causally separable
spacelike strings $\mathcal{S}=x+\mathbb{R}_{+}e$ which together with their
scalar sl "escorts" $\phi(x,e)$ are linearly related to their pl counterpart
simple illustration is provided by the interaction density $L^{P}=A_{\mu}%
^{P}j^{\mu}$ of massive QED which is related to its pl counterpart as $A_{\mu
}(x,e)=A_{\mu}^{P}(x)+\partial_{\mu}\phi(x,e)$.$~$

Its use in an interaction density of e.g. massive QED $L^{P}=A_{\mu}^{P}%
j^{\mu}$ results in a relation $L^{P}=L-\partial^{\mu}\phi j_{\mu}$ in which
the sl density $L(x,e)~$has an improved short distance dimension $d_{sd}(L)=4$
(instead of $d_{sd}(L^{P})=5$) and accounts for the first order contribution
to the (on-shell) $S$-matrix in the adiabatic limit $S=\int L$ to which the
boundary term from $V_{\mu}=\phi j_{\mu}$ does not contribute.

This is in a nut-shell a perturbative implementation of the aforementioned
abstract Buchholz-Fredenhagen theorem; it secures the existence of
interpolating sl fields whose directional smearing provides the B-F operators
localized in arbitrary narrow spacelike causally separable cones and insures
that their large-time scattering limits results in $e$-independent Wigner
particles and their S-matrix.

The extension of this first order $L$ to higher orders involves time-ordered
products in the interaction densities $L~$respective $L^{P}$ and leads to new
powerful normalization conditions which insure that the two different
interaction densities lead to the same S-matrix. As a result of the with
perturbative order growing number of counterterms, the $L^{P}$ theory by
itself is physically useless; but being "guided" by the sl $L,V_{\mu}$ pair it
becomes a well-defined physically useful companion which shares not only its
parameters but also its S-matrix and local observables with the $d_{sd}(L)=4$
SLFT$.$ Its only memory about its "unguided past" is the with perturbative
order increasingly singular $d_{sd}\rightarrow\infty$ short distance dimension
of its interpolating fields.

\textit{SLFT is an S-matrix theory in the sense that the particle content
together with the string-independence of }$S$\textit{ determines (in all cases
studied up to now uniquely) the form of the interaction density. In a second
step the construction of the S-matrix is extended to that of pl and sl
interacting fields. }

Different from Lagrangian quantization the SLFT formalism does not prefer
certain fields. All interacting fields which act in the same Wigner-Fock space
and are members of the same causality class are on equal footing; which
particle they interpolate depends only on the nontriviality on their
vacuum-to-one-particle matrix elements.

Often new theoretical insights are the result of accidental observations. SLFT
is not of this kind; what led to it is the rather deep connection of sl
localization with \textit{modular localization} theory. The terminology
"string" used in quantizations of classical actions (Nambu-Goto actions,
world-sheets,..) bears no relation to the causal localization of string-local
quantum fields in the present work.

A definition of causal localization which avoids such misunderstandings is
that in terms of \textit{modular localization. }In fact modular localization
permits to identify\textit{ a pre-form of causal localization already within
the Wigner positive energy representation space} \cite{BGL} before "second
quantization" converts it into the algebraic form of Einstein causality in
QFT. \textit{This idea paved the way for the construction of the QFT behind
Wigner's infinite spin representation}.

Modular localization theory can be traced back to the Tomita-Takesaki modular
theory of operator algebras of the 60's. It is one of a few mathematical
theories to which physicists working on problems of statistical mechanics of
open systems \cite{HHW} made important contributions. It made its first
appearance in the context with causal localization in the Bisognano-Wichmann
theorem \cite{BW} which deals with modular properties of wedge-localized
algebras. \textit{Modular operator theory and modular localization requires
positivity and hence cannot be applied to GT and Lagrangian quantization}.

As the result of accommodating thermal aspects and causal localization under
one conceptual roof, it led to profound insights (thermal properties of "event
horizons") into Hawking's black hole radiation \cite{Sew}. A first survey
about its history enriched by new results was presented by Borchers \cite{Bo}.

Modular localization also played an important role in the construction of QFTs
from S-matrices of integrable models in $d=1+1$ dimensions \cite{AOP}
\cite{AL}. Presently ideas from modular operator theory are being successfully
applied to obtain a foundational understanding of entanglement entropy caused
by causal localization (for a survey see \cite{Wi} \cite{Ho-Sa} and references therein).

Modular localization theory permits to extend Weinberg's intertwiner
construction to Wigner's infinite spin representations and to obtain explicit
expressions for the associated sl fields \cite{MSY1,MSY2,Koe}\cite{PY}. More
recently these fields reappeared as "Pauli-Lubanski limits" of finite spin sl
fields \cite{PL}. This made it possible to investigate physical properties of
quantum matter through the study of its positivity-obeying causal localization
structure and look for theoretical reasons why certain types of matter can not
be seen in counters \cite{dark}.

One should also mention a series of more recent publications \cite{Sch-T}
\cite{Bekaert} in which covariant wave functions were constructed, but the
much stronger result in the aforementioned work was overlooked. Relativistic
wave equations for infinite spin appeared already in Wigner's 1948 paper
(\cite{Wig 48} 12.1-12.4). These different wave functions describe different
covariant bases in Wigner's irreducible representation space. But for studying
physical manifestations of matter one needs to know its causal localizability
which in case of infinite spin does not follow from covariance and needs the
use of modular localization theory as used in the cited 2006 papers. In this
way the 1970 No-Go theorem \cite{Y} which excluded point-like localization was
replaced by a sl Yes-Go theorem.

In fact the exclusion of linear pl fields is part of a more general No-Go
theorem which rules out the possibility of constructing pl composites from the
linear sl infinite spin fields. In its most general form the theorem excludes
the existence of operator algebras localized in finite spacetime regions
\cite{LMR}.

These theorems against pl localization of infinite spin matter may be seen as
an extreme counterpart of the Weinberg-Witten No-Go theorem against the
existence of higher helicity conserved currents and energy-momentum tensors.
The difference is that there still exist W-W local charges, whereas in case of
infinite spin there are no nontrivial operators localized in finite regions.

Recall that the raison d'\^{e}tre of a relativistic quantum \textit{field}
theory (for the difference between QFT and relativistic QM, see section 3 in
\cite{SHPMP}) is the realization of the "Nahewirkungsprinzip" (action in the
neighborhood principle) of Faraday and Maxwell which culminated in Einstein's
concept of relativistic covariance and causal localization. The positivity
requirement of quantum probability turns the construction of models of QFT
into a challenging problem which gauge theory did not solve.

It is the aim of this work to show how the recent SLFT formulation solves
problems which have remained outside the range of GT (for a review of such
problems see \cite{Ra-Ta}). In \cite{MRS1} \cite{MRS2} this was already
achieved for the problem behind the Weinberg-Witten (W-W) No-Go theorem
\cite{W-W} and the $s=2$ van Dam-Veltman-Zakharov (D-V-Z)\ discontinuity
\cite{DVZ,Z}. Here we add the causality problems raised by Velo and Zwanziger
(V-Z) \cite{V-Z}.

SLFT's central point is however the presentation of a sl-based perturbation
theory which in contrast to gauge theory preserves the Hilbert space
positivity (no indefinite metric- and ghost-degrees of freedom) without
destroying the causal separability of fields. It leads in particular to
interesting different physical interpretations of interactions between vector
mesons and Hermitian fields (Higgs models, but no SSB Higgs mechanism)
\cite{Bey}.

The content of this paper is organized as follows.

The next section recalls and extends recent (partially already published)
results concerning the construction of causally separable string-local free
fields. It consists of 4 subsections which includes the construction of sl
massless vector potentials and their canonically related massive counterpart.

The third section addresses the problem of interactions with external
potentials. It is shown that the origin of the Velo-Zwanziger causality
problem is the incorrect expectation that by modifying free field equations by
adding linear couplings to external potential one preserves causality in the
sense of causal propagation of Cauchy data. The solution of the V-Z problem
has a close formal proximity to the solution of the Weinberg-Witten problem in
\cite{MRS2}.

Section 4 provides some background about modular localization. Its aim is to
show that causal localization is incompatible with any form of quantization
but important for understanding properties of causally localized quantum matter.

In section 5 the SLFT renormalization theory is applied to calculation of the
S-matrix in various models involving vector mesons. including some speculative
remarks on $s\geq2$ interactions.

Section 6 addresses problems of interacting sl fields in particular the
model-dependent distinction between pl observables and sl interpolating fields.

The concluding remarks in section 7 summarize the new insights and present an outlook.

\section{String-local tensor potentials and conservation laws}

\textit{This section provides the kinematical prerequisites of SLFT i.e. the
construction of those sl free fields which are used in later sections for the
calculation of the S-matrix and interacting fields. The kinematic localization
of free fields is not the same as the dynamic localization of their
interacting counterparts (section 6).}

\subsection{Massless string-local potentials}

The fact that even in the absence of interactions massless gauge potentials
have no positivity-maintaining pl counterpart led to a more foundational
re-thinking regarding the relation between positivity and causal
localizability for which the solution of the massless infinite spin problem in
terms of sl fields served as a role-model \cite{MSY1,MSY2}. The finite
helicity problem is simpler since in this case their exists only one covariant
family of sl potentials $\hat{A}_{\mu}$ in the Wigner-Fock helicity Hilbert
space whose field strength is the pl field strengths
\begin{equation}
\partial_{\mu}\hat{A}_{\nu}(x)-\partial_{\nu}\hat{A}_{\mu}(x)=F_{\mu\nu}(x)~~
\label{1}%
\end{equation}

They have the form of a semi-infinite line integrals (strings, rays)
\begin{align}
&  \hat{A}_{\mu}(x)=A_{\mu}(x,e):=\int F_{\mu\nu}(x+\lambda e)e^{\nu}%
=:(I_{e}F_{\mu\nu})(x)e^{\nu}\label{2}\\
&  U(a,\Lambda)A_{\mu}(x,e)U(a,\Lambda)^{\ast}=(\Lambda^{-1})_{\mu}^{\nu
}A_{\nu}(\Lambda x,\Lambda e)~\nonumber
\end{align}
with $e$ representing a space- time- or lightlike vector \textit{which
participate in the transformation under the homogenous Lorentz group}. Causal
localizability requires the possibility of placing an arbitrary large number
of such sl fields in relative spacelike separated positions (denoted as
${\lower1pt\hbox{\LARGE$\times$}}$ ). This excludes the timelike case but
permits space- and light-like strings\footnote{This is less obvious in the
lightlike case.}%

\begin{equation}
\left[  A_{\mu}(x,e),A_{\mu}(x^{\prime},e^{\prime})\right]  =0,~x+\mathbb{R}%
_{0}^{+}e\,{\lower1pt\hbox{\LARGE$\times$}}\,x^{\prime}+\mathbb{R}_{0}%
^{+}e^{\prime}%
\end{equation}
Spacelike unit vectors $e$ with $e^{2}=-1~$are points on the $d=1+2$ unit de
Sitter space, whereas lightlike vectors $e$ with $e^{2}=0~$may be identified
with points on the two-dimensional celestial sphere. A closer examination
shows that line integrals of \textit{massless} field strengths along lightlike
lines are ill-defined (see below) but well-defined (as distributions in $e$)
for spacelike $e$; time-like lines would violate causal separability.

The derivation of nonperturbative theorems (PCT, Spin\&Statistics, cluster
properties, LSZ scattering theory,..) does not need pl fields; what is
important is the preservation of causal separability i.e. the fact that one
can place an \textit{arbitrary number} of sl fields into relative spacelike position.

The mathematical status of sl fields requires a more careful look at their
singularity structure. For this purpose it is convenient to compute their
2-point function (2-pfct). Starting from that of the field strengths
\begin{align}
\left\langle F_{\mu\nu}(x)F_{\kappa\lambda}(x^{\prime})\right\rangle  &  =\int
e^{-ip(x-x^{\prime})}M^{F_{\mu\nu},F_{\kappa\lambda}}(p)d\mu_{0}(p),\text{
\ }d\mu_{m}(p)=\frac{d^{3}p}{2\sqrt{\vec{p}^{2}+m^{2}}}\\
M^{F_{\mu\nu},F_{\kappa\lambda}}(p)  &  =-p_{\mu}p_{\kappa}g_{\nu\lambda
}+p_{\mu}p_{\lambda}g_{\nu\kappa}-p_{\nu}p_{\kappa}g_{\mu\lambda}+p_{\nu
}p_{\lambda}g_{\mu\kappa}\nonumber
\end{align}
and, using the fact that the $\lambda$-integration amounts to the Fourier
transform of the Heavyside function and hence leads to distribution
$(pe)_{i\varepsilon}^{-1}=\lim_{\varepsilon\rightarrow0}(pe+i\varepsilon
)^{-1}$ as boundary values of analytic functions, one obtains \cite{MRS2}%
\begin{align}
&  \left\langle A_{\mu}(x,-e)A_{\nu}(x^{\prime},e^{\prime})\right\rangle =\int
e^{-ip(x-x^{\prime})}M^{A_{\mu},A_{\nu}}(p,e,e^{\prime})d\mu_{0}(p)\label{3}\\
&  M^{A_{\mu},A_{\nu}}(p,e,e^{\prime})=E_{\mu\nu}(-e,e^{\prime})=-\eta_{\mu
\nu}+\frac{p_{\mu}e_{\nu}}{(pe)_{i\varepsilon}}+\frac{e_{\mu}^{\prime}p_{\nu}%
}{(pe^{\prime})_{i\varepsilon}}-\frac{(ee^{\prime})p_{\mu}p_{\nu}%
}{(pe)_{i\varepsilon}(pe^{\prime})_{i\varepsilon}}\nonumber
\end{align}
where the tensor $E_{\mu\nu}$ turns out to be an important building block of
higher helicity 2-pfcts$.$ The scaling degree $d_{sd~}$ is defined as the
leading short distance contribution $\lambda^{-2d_{sd}\text{ }}$of the 2-ptfct
under the scaling $\xi\rightarrow\lambda\xi,$ $\xi=x-x^{\prime}$ for
$\lambda\rightarrow0~$and can be directly read off from the large momentum
behavior. Whereas $d_{sd}(F)=2,$ the line integration lowers the degree to
$d_{sd}(A)=1$.

A more detailed study shows that sl potentials and their 2-pfcts are
well-defined as \textit{distributions} in $e,~e^{2}=-1$ (the unit de Sitter
space) and $x.$ All operators and correlation function are of homogeneous
degree zero and hence the de Sitter differential can be written in the
covariant form $d_{e}=de_{\mu}\frac{\partial}{\partial e_{\mu}}.$ For
lightlike $e^{\prime}s$ and $m>0$ the last term in (\ref{3}) vanishes for
$e^{\prime}=$ $-e$ and the distributional dependence of $A_{\mu}(x,e)~$on $e$
changes to that of a function so that a directional testfunction smearing in
$e\ $is not necessary. The identification of $e^{\prime}s$ in products of
fields leads to a significant notational simplifications in perturbative
calculations. The existence of momenta for which $p$ is parallel to $e$
excludes however massless limits lightlike strings.

For the timelike directions the denominators never vanish and no smearing is
needed, but the causality requirement, namely the existence of an arbitrary
number of causally separated sl fields, cannot be satisfied. Hence the choice
$e=e_{0}=(1,0,0,0)$ leads to the nonlocal Coulomb- (or radiation-) potential
with $A_{0}^{C}=0$ and spatial components%
\begin{equation}
M^{A_{i}^{C}A_{j}^{C}}=\delta_{i,j}-\frac{p_{i}p_{j}}{\mathbf{p}^{2}}%
\end{equation}
"Freezing" this timelike string direction destroys the covariant
transformation and one obtains a noncovariant inhomogeneous transformation law
in which only the rotations and translations maintain their covariant
appearance (see \ref{inh} below). Full covariance can be restored by letting
the timelike direction participate in the Lorentz transformation, but the loss
of causal localization remains. The Coulomb potential is used in quantum
mechanics where relativistic covariance and causality play no role.

It is interesting to note that the Coulomb potential results also from
\textit{averaging a spacelike string over spatial directions in the }%
$t=0$\textit{ plane orthogonal to the timelike }$e_{0}~$\textit{vector}. There
is no direct way to undo this directional averaging; one rather has to return
from $A^{C}$ to its covariant field strength $F$ and obtain the associated sl
potential as in (\ref{2}). This directional averaging reveals a \textit{close
formal connection between the axial- and Coulomb- "gauge"}. Both potentials
exist in the same Wigner-Fock helicity space, but only the covariant sl
potential (\ref{2}) is manifestly causal.

The use of sl potentials turns the so-called noncovariant axial- and
lightcone-gauges into better manageable covariant Einstein-causal fields which
act in a positivity maintaining Hilbert space.

It should be mentioned that in the literature the terminology "gauge" is used
with two different meanings. In the covariant setting of QED perturbation
theory it refers to a formal symmetry whose generator is a "gauge charge"
which depends unphysical indefinite metric degrees of freedom. On the other
hand the Coulomb- or axial- gauge contains only the two helicity $h=\pm1$
degrees of freedom and there is no symmetry-implementing gauge charge,
although the additive contribution to the Lorentz transformation looks like a
non-covariant gauge transformation (\ref{inh}) re-expressing the
Lorentz-transformed $e=\Lambda e_{0}$ in terms of original $e_{0}$.

It is not the aim of this work to change historically grown terminology. Here
the terminology "gauge" is exclusively used the situation in which unphysical
degrees of freedom provide a covariant "gauge symmetry". Quantum gauge
symmetry is not a physical symmetry (and consequently there is no physical
sense in which it can be broken) but rather a formal tool to extract a
physical theory as a subtheory from an unphysical formalism.

The large momentum behavior of the 2-pfct determines the short distance
behavior of the field whereas the distributional behavior in $e$ depends on
the dimensionality of spacelike $e$-directions on which $pe$ vanishes. The
case of \textit{lightlike} $e^{\prime}s$ is a bit more tricky. For massive $p$
the $pe$ denominator does not vanish since $p$ and $e$ only touch at lightlike
infinity and as a result the sl fields are functions in $e$.

This changes in a radical way for massless $p;$ in that case for each $e$
there are lightlike $p^{\prime}s$ on which $pe$ vanishes and as a result
massless fields localized on lightlike strings do not even exist in the sense
of distributions \footnote{I am indebted to Henning Rehren for drawing
attention to the nonexistence of massless lightlike string localized fields.}.
Lightlike sl fields have an interesting connection with light-cone
quantization. In the massless case they reveal in a much clearer way the
problematic nature of "lightcone quantization" \cite{Leib}.

The main purpose of this work is to offer a positivity- and
causality-\ preserving alternative to gauge theory which avoids the use of the
quantization parallelism to classical field theories by starting from Wigner's
manifestly positivity-preserving particle representation theory. The important
point is that spacetime localization properties already exist in the pre-form
of \textit{modular localization} within Wigner's particle theory. They can be
used to construct pl or sl intertwiner functions which convert Wigner's
creation and annihilation operators into covariant pl or sl free fields.

The perturbative construction of the S-matrix and of interacting sl fields
does not need modular localization theory. For problems as
\textit{localization entropy \cite{Ho-Sa} \cite{Wi}} and nonperturbative
constructions \cite{AL} its use is however indispensable. In the context of
the present paper its importance is based on pinning causal localization to
quantum positivity; whereas Lagrangian quantization of fields allows the
presence of unphysical degrees of freedom, modular localization excludes them.
More remarks on modular localization will be deferred to section 4.

The construction of sl potentials in terms of pl field strengths (\ref{2})
permits an iteration to a scalar potential\ $\Phi$%
\begin{equation}
A_{\mu}(x,e)-A_{\mu}(x,e^{\prime})=\partial_{\mu}\Phi(x,e,e^{\prime}%
),~\Phi=(I_{e^{\prime}}I_{e}F_{\mu\nu})(x)e^{\mu}e^{\prime\nu}~ \label{5}%
\end{equation}
The $\Phi$ represents a field which is localized on the 2-dimensional
\textit{conic region} $\lambda e+\lambda^{\prime}e^{\prime},~\lambda
,\lambda^{\prime}\geq0.$ In the massless limit this flux $\Phi$ is
logarithmically divergent. The logarithmic divergence is expected to lead to
an $e,e^{\prime}$ dependent continuous set of superselection rules which
extend the Wigner-Fock helicity space.

This is reminiscent of the behavior of the exponential of a massive scalar
free field in $d=1+1$ in the massless limit\footnote{This infrared behavior
was first observed in the coupling of a $d=1+1$ current to the derivative of a
massless scalar field ("infraparticle" \cite{Infra} ).} which played an
important role in the work on "bosonization" of massless fermions and anyons
\cite{S-S}. In that case the massless limit of the properly mass-normalized
exponentials leads to the superselection property%

\begin{equation}
\left\langle e^{ia_{1}\varphi(x_{1})}\dots e^{ia_{n}\varphi(x_{n}%
)}\right\rangle =0\quad\hbox{if}~~\sum_{1}^{n}a_{i}\neq0. \label{scal}%
\end{equation}
corresponding to $a_{i}$-"charge" conservation.

The "photon cloud" in the $e$-direction associated with $\exp ig\varphi$ is
expected to cause a directional superselection rule which appears in the form
of $e^{ig\varphi}\psi$ in the large time behavior of electric charge carrying
fields and causes the modification of LSZ scattering theory. In this way one
may hope to obtain a genuine spacetime understanding of the infrared momentum
space recipes in \cite{YFS}.

The interest in this problem is also motivated by the existence of rigorous
results derived from an appropriate formulation of the quantum Gauss law
\cite{Bu}. This theorem states that interacting electric charge-carrying
operators $\psi$ are accompanied by spacelike extended "photon clouds"$~$whose
different asymptotic conic directions correspond to a continuum of
superselection sectors within the same charge-carrying sector. This is the
cause a spontaneous breaking of Lorentz symmetry \cite{Froh}.

The existence of a continuum of superselection sectors for free photons would
suggest the existence of large time asymptotic charge-carrying matter fields
of the form $\psi_{0}e^{ig\Phi}$ with $\psi_{0}$ a free matter field. Their
large time asymptotic behavior is expected to play an important role in a
future spacetime understanding of infrared properties which is outside the
physical range of gauge theories.

For many applications it is useful to encode change in $e$ (\ref{5}) into
changes of the Lorentz transformation law. A differential relation which is
the basis for such conversion has the form (\cite{MRS2} Corollary
3.3)\footnote{As mentioned therein this remains well defined since sl fields
and their correlation functions are homogeneous functions of degree zero in
$e$ and $p.$}
\begin{equation}
d_{e}A_{\mu}(x,e)=\partial_{\mu}u(x,e),\ ~d_{e}=\sum_{i}d_{e_{i}}%
\partial^{e_{i}} \label{6}%
\end{equation}
where $u$ is an exact de Sitter one-form $u=d_{e}\phi$. This conversion of
directional de Sitter differentials into $x$-derivatives plays an important
role in passing from interactions in the presence of a mass gap to their
massless limit.

In the present context the formula for the change of $e^{\prime}s$ can be used
to compute the additive change which is necessary in order to
\textit{maintain} the timelike $e_{0}$ direction of the Coulomb potential
$A_{i}^{C},~A_{0}^{C}=0$. The resulting affine transformation formula
\begin{equation}
U(a,\Lambda)A_{i}^{C}(x)U(a,\Lambda)^{\ast}=(\Lambda^{-1})_{i}^{~l}A_{l}%
^{C}(\Lambda x+a)+(\Lambda^{-1})_{i}^{~\mu}\partial_{\mu}\chi(x) \label{inh}%
\end{equation}
is equivalent to that obtained by starting from the Wigner helicity
representation and using transverse polarization vectors \cite{Wein}.

A similar situation arises if one fixes an "axial" direction as
e.g.\ $e=(0,1,0,0).$ In this case the causal localizability is preserved in
both descriptions. Ignoring the spacetime localization aspect and treating the
axial direction as a noncovariant gauge misses the necessity of directional
smearing (smearing around a point in de Sitter space) and probably contributed
to the abandonment of the "axial gauge fixing". But what became a curse in the
axial gauge fixing turns out to be a blessing in the covariant SLFT setting.

Covariant gauges as used in covariant perturbation theory always require the
presence of ghost-extended indefinite metric BRST degrees of freedom setting
which reduces the physical range. SLFT cuts the umbilical cord between
perturbative Lagrangian quantization and classical gauge theory and restores positivity.

\subsection{A brief interlude, relation with concepts of algebraic QFT}

The simplest illustration of the interplay between positivity and causality is
provided by the Aharonov-Bohm effect. To see this recall that Einstein
causality is the statement that the algebra of operators localized in the
causal complement $\mathcal{O}^{\prime}$ of a spacetime region $\mathcal{O}$
belong to the commutant $\mathcal{A}(\mathcal{O})^{\prime}$ algebra (the von
Neumann algebra which consists of all operators which commute with
$\mathcal{A}(\mathcal{O})$)
\begin{align}
\mathcal{A}(\mathcal{O}^{\prime})  &  \subseteq\mathcal{A}(\mathcal{O}%
)^{\prime}\quad\hbox{or}\quad\mathcal{A}(\mathcal{O})\subseteq\mathcal{A}%
(\mathcal{O}^{\prime})^{\prime},\quad\hbox{Einstein causality}\label{Haag}\\
\mathcal{A}(\mathcal{O}^{\prime})  &  =\mathcal{A}(\mathcal{O})^{\prime}%
,\quad\hbox{Haag duality}\nonumber
\end{align}
The second line defines the somewhat stronger Haag duality which states that
an operator which commutes with all operators localized in the causal
complement of $\mathcal{O}~$\textit{must} belongs to $\mathcal{A}%
(\mathcal{O})$.

Einstein causality is a defining property of relativistic QFT, but Haag
duality may be violated. In the absence of interactions such a violation can
be excluded for massive QFT's but it does occur in the massless case when the
$2s+1$ spin degrees of freedom are converted into the $\pm h$ helicities. As
observed in \cite{LRT} (unpublished) Haag duality, which holds for simply
connected spacetime regions, is violated for multiply connected regions as
(genus one) tori.

In their proof the authors carefully avoid the use of gauge potentials.
associated to $m>0$ massive is a property ($g\geq1~$tori). This violation is
an intrinsic property of the operator algebra generated by the field strength
$F_{\mu\nu}~$of the $h=1~$Wigner representation. But if one wants to
understand this in terms of vector potentials one must use the
positivity-maintaining sl potentials which preserve a somewhat hidden
topological properties of Wilson loops which cause the breakdown of Haag
duality while it upholds Einstein causality \cite{Bey}. The indefinite metric
potentials cannot distinguish between the two; only the localization in the
presence of positivity is physical.

The cause of "eeriness" about the Aharonov-Bohm effect \cite{A-B}~(but also of
its popularity) is that we erroneously interpret the intuitively accessible
geometric Haag duality with the more abstract Einstein causality, thus
forgetting that the latter also admits operators which have no unambiguous
causal localization region (e.g. the magnetic flux through a surface with a
fixed boundary). The ideal solenoid in the A-B setup closes at spacelike
infinity, which in the conformal Wigner-Fock helicity world is a circle. In
case of a finite tube one must place the electric circuit into a region of
little magnetic backflow from north- to south-pole.

This is a strong reminder that it is not possible to separate causality from
positivity and a warning not to confuse the "fake localization" of gauge
dependent objects with genuine causal localization of quantum matter. It
points to a potential source of misunderstanding involved in transferring the
perfectly reasonable classical notion of \textit{local} gauge symmetries to
QFT by attributing a physical meaning to the formal observation that quantum
gauge charges are "more local" than those corresponding to internal
symmetries. It is also a reminder to rethink the physical meaning behind the
terminology "gauging a model".

From (\ref{6}) it follows that a Wilson loop\footnote{By convoluting with a
test function one can convert the Wilson loop integral into into an operator
localized on a solid torus.} formed with $A_{\mu}(x,e)$ \textit{is independent
of} the$~$choice$~$of the direction $e~$\cite{Bey}$.~$However it
\textit{retains a topological memory} of the string directions of the
integrand which prevents a naive materialistic identification with a
localization in a torus. One can choose $e^{\prime}s$ in such a way that this
extension is spacelike with respect to any simply connected convex compact
region\footnote{In case the solenoid has open ends the Wilson loop should
avoid the region of the north-south magnetic backflow.}. Yet it is not
possible to completely forget that the vector potential has a directional
$e$-dependence. An elegant formulation of this $h\geq1~$topological phenomenon
directly based on field strengths and their duals in terms of "linking
numbers" can be found in \cite{BCRV}.

\subsection{Massive string-local potentials}

Before passing to the construction of massive sl fields it is helpful to
recall the construction of their pl intertwiner functions $v(p)$ which convert
the$~m>0\ $Wigner creation and annihilation operators $a^{\#}(p,s)~$into
covariant \cite{Wein}. For the $s=1$ Proca field they are the three
polarization vectors~$v_{\mu}(p,s_{3})$ obtained by applying a rotation-free
Lorentz boost to the spatial coordinate unit vectors. By definition they are
Minkowski-orthogonal to $p_{\mu}$ and hence correspond to the 3 polarization
vectors $v$ obeying the completeness relation
\begin{align*}
&  \,\sum_{s_{3}=--1}^{1}v_{\mu}(p,s_{3})v_{\nu}(p,s_{3})=-\eta_{\mu\nu}%
+\frac{p_{\mu}p_{\nu}}{m^{2}}\\
&  M^{A_{\mu}^{P},A_{\nu}^{P}}(p)=-\pi_{\mu\nu}(p),\quad\pi_{\mu\nu}%
(p)=\eta_{\mu\nu}-\frac{p_{\mu}p_{\nu}}{m^{2}}%
\end{align*}
where the $\pi_{\mu\nu}$ of the momentum space 2-pfct which also turns out to
be the basic building block of all higher spin massive tensor potentials.

With a pl Proca potential $A_{\mu}^{p}~$one may associate two sl fields, the
scalar sl field $\phi~$defined in terms of a line integral $I_{e}$ of the
$e$-projected Proca field $A_{\mu}^{P}e^{\mu}$ along $e$ starting from the
point $x$
\begin{equation}
\phi(x,e)=(I_{e}A_{\mu}^{P})(x)e^{\mu},~~a(x,e)=-m\phi(x,e) \label{a}%
\end{equation}
and the sl vector potential $A_{\mu}$ in terms of the field strength of the
Proca potential%

\begin{equation}
A_{\mu}(x,e)=(I_{e}F_{\mu\nu})(x)e^{v},\text{ }F_{\mu\nu}(x)=\partial_{\mu
}A_{\nu}^{P}-\partial_{\nu}A_{\mu}^{P} \label{F}%
\end{equation}
whose massless limit coincides with (\ref{2}).

The multiplication with $m$ in (\ref{a}) restores $d_{cl}=1$ and removes the
mass singularity, so that the $m=0$ limit is an $e$-independent massless
scalar $d_{sd}(a)=1$ free field. In fact for all massive or massless sl tensor
potentials $d_{sd}=1$ and $3/2$ for halfinteger $s$ whereas their pl
counterparts increase linearly as $d_{sd}=s+1~$or $d_{sd}=s$ and diverge like
$m^{-s}$ for $m\rightarrow0$

In particular the momentum space 2-pfct of the massive field strength and its
massless associated sl vector potential are identical to their massless
counterparts. This permits to lower the number of degrees of freedom by
passing from $p\in H_{m}^{\uparrow}$ to$~p\in V^{\uparrow}$ (and its
"fattening" inversion, see below)$.$ In the pl setting this is not possible or
can only be achieved in the presence of indefinite metric degree of freedom
(the DVZ discontinuity, the WW problem).

It is instructive to look at this degrees of freedom balance in more detail
\cite{MRS1,MRS2}. With the help of a $p$-dependent 4-matrix $J$ (complex
conjugation changes the sign of $e$,~$tr~$=~transposed)
\begin{align}
&  J_{\mu}^{~~\nu}(p,e)=\eta_{\mu}^{\ \ \nu}-\frac{p_{\mu}e^{\nu}%
}{(pe)_{i\varepsilon}},~\overline{J(p,-e)}=J(p,e)\\
&  M^{A_{\mu}(-e),A_{\nu}(e)}=:E_{\mu\nu}(e,e)=(J\pi J^{tr})_{\mu\nu}\nonumber
\end{align}
the in $e$ diagonal momentum space 2-pfct takes the form of the second
line\footnote{Taking the same $e$ would lead to the distributionally
ill-defined denominator $pe_{-i\varepsilon}pe_{i\varepsilon}.$}. It shows that
the positivity of the sl 2-pfct is inherited from the pl positivity. The rank
of the $E$-matrix accounts for the degrees of freedom is 3 as a result of
$J^{tr}e=0$ and the additional relation $E_{\mu\nu}p^{\nu}=0$ for $p\in
V^{+}~$leads to a reduction from the three spin component to the two
helicities $h=\pm1$. This degrees of freedom counting breaks down in the
presence of indefinite metric.

This descend from $p\in H_{m}^{\uparrow}$ to $V^{+}$ permits an inversion,
namely by continuous passing from momenta $p\in V^{+}$ to the mass shell
$H_{m}^{+}$~("fattening") one creates a new physical degree of freedom which
together with former $\pm1$ accounts for the $3$ degrees of freedom of spin
$s=1$. Such "magical" conversion of the particle content of two inequivalent
Wigner representations can neither be achieved in terms of pl fields (no
massless limit) or become contaminated by the presence of indefinite metric
causing and ghost degrees of freedom of gauge theory. This is of particular
interest in case of $s=2$ \cite{MRS1,MRS2} (see below). The use of sl fields
is even more important for passing to the massless limit in the presence of
interactions involving higher spins.

In the literature the terminology "fattening" had been used in connection with
the Higgs model which describes the interaction between a massive vector meson
with a massive real scalar field $H$ as the result of spontaneous breaking of
gauge symmetry (the Mexican hat potential). This idea contains two conceptual
misunderstandings (which will be commented on in section 5 and the concluding remarks).

The real power of SLFT emerges in models of \textit{selfinteracting} massive
vector mesons where the preservation of 2nd order renormalizability requires
the compensatory presence of a coupling to a Hermitian scalar $H$ (Higgs)
field\footnote{This (and not SSB) is the \textit{raison d'\^{e}tre for the
}$H$ (section 5).} and imposes a Lie-algebra structure on the leading terms in
the $A_{\mu}~$self-interactions. In section 5 we will provide the arguments.

An important property of the previously introduced pl and sl vector potential
and its scalar escort $\phi(x,e)$ is their linear relation
\begin{equation}
A_{\mu}(x,e)=A_{\mu}^{P}(x)+\partial_{\mu}\phi,\quad\phi=-\frac{1}{m}a
\label{escort}%
\end{equation}
This property justifies to call the $\phi^{\prime}s$ "escorts" of the sl
potential, they share the same degrees of freedom. The appearance of the
escort in form of a derivative is a consequence of Poincar\'{e}'s lemma. The
linear relation between fields corresponds to that between intertwiners ($J$
as before):%
\begin{equation}
J_{\mu}^{\ \ \nu}v_{\nu}(p)=v_{\nu}-p_{\mu}\frac{(ve)}{(pe)_{i\varepsilon}}%
\end{equation}
which follows directly from the definition (\ref{F}). Each field contains the
full information of the ($m,s=1$)~Wigner representation; the encoding of $s=1$
into a scalar is only possible within sl.

It is not accidental that the massive vector potentials which result from
"fattening" their unique massless counterpart play a distinguished role in the
new SLFT renormalization theory. Their smooth connection represents the higher
spin analog of the smooth relation between $s<1$ massless fields and their
massive counterpart. The weakening of localization is necessary to preserve
this smoothness in the presence of change of the number of degrees of freedom.

In the massless limit the $A_{\mu}^{P}(x)$ and $\phi(x,e)$ diverge as
$m^{-1}~$whereas the $A_{\mu}(x,e)$ and $a(x,e)$ stay infrared finite. The
relation ($u~$was introduced in (\ref{6}))%
\begin{align}
\partial^{\mu}A_{\mu}  &  =-ma,~~d_{e}A_{\mu}=\partial_{\mu}u\label{c}\\
u  &  =-m^{-1}d_{e}a
\end{align}
leads to a divergence-free massless vector potential (Lorentz
condition\footnote{Note that this is an operator identity and not an imposed
gauge condition.})~and a relation between two massless 1-forms in the de
Sitter space of spacelike directions (that which remains of (\ref{escort})).
The purpose of the mass factors is to preserve the relation $d_{sd}=d_{cl}%
~$for all sl fields. The massless limit of $u$ is logarithmically divergent.

Escorts (whose number increase with $s$) do not contain new degrees of freedom
since, as the pl $A^{P},~$they are linear in the Wigner $s=1$
creation/annihilation operators $a^{\#}(p,s_{3})$ and only differ in their
intertwiners. Rearrangements of degrees of freedom are quite common in quantum
mechanical many-body problems\footnote{A well-known case is the appearance of
Cooper pairs encounters in passing to the low temperature superconducting
phase. Without this rearrangement classical vector potentials would not become
short range inside a superconductor (the London effect).}. Escorts are
rearranged $s=1$ degrees of freedom which carry the full content of ($m>0,s$)
Wigner representations.

For $s\geq2$ the sl fields lead to new properties. As a result of a possible
relation with gravitation the case $s=2~$is of special interest. The
intertwiner of spin $s$ Proca potentials (the $P$ in $A_{...}^{P}$ refers to
Proca or alternatively to pointlike) must be a divergence- and trace-free
symmetric tensor; this is a consequence of the way the $2s+1$ component
subspace of spin is embedded in the $3s$-fold tensor product. Hence the
intertwiners $v_{\mu_{1}\dots\mu_{s}}(p,s_{3})\ $convert the symmetric
trace-free $s$-fold tensor product of three-component spin $1~$polarization
vectors into covariant tensors of tensor-degree $s.$

For the momentum space $s=2~~$2-pfct one obtains
\begin{equation}
M^{A_{\mu\nu}^{P},A_{\kappa\lambda}^{P}~}(p)=\frac{1}{2}\left[  \pi_{\mu
\kappa}\pi_{\nu\lambda}+\pi_{\mu\lambda}\pi_{\nu\kappa}\right]  -\frac{1}%
{3}\pi_{\mu\nu}\pi_{\kappa\lambda}%
\end{equation}
where the numerical factors have their combinatorial origin in the symmetry
and tracelessness and hence depend on the degrees of freedom. The sl 2-pfcts
are of the same algebraic form and result by substituting $\pi_{\mu\nu
}\rightarrow E_{\mu\nu}(-e,e)$ \cite{MRS1,MRS2}. As for $s=1~$this can be seen
by passing from the Proca potential to the field strength ($as$ stands for
antisymmetrisation)
\begin{equation}
F_{\mu_{1}\nu_{1}\mu_{2}\nu_{2}}=\underset{\mu\leftrightarrow\nu}{as}%
~\partial_{\mu_{1}}\partial_{\mu_{2}}A_{\nu_{1}\nu_{2}}^{P}%
\end{equation}
and using the two-fold momentum space $I$ operation to pass from the field
strength to the potentials. Note that the symmetry of the Proca potential
reduces the anti-symmetrization to a pairwise operation $\mu_{i}%
\leftrightarrow\nu_{i}.$ The resulting permutation properties of the resulting
$F$ are those of the linearized Riemann tensor.

The new phenomenon for $s>1$ is that the massless limit of this field strength
is not the same as that obtained directly from the massless $h=\pm2$ Wigner
representation. Correspondingly the sl potential associated with the massless
limit of $F^{s=2}$ is different from that of$~F^{\left\vert h\right\vert =2}$
\begin{align}
A_{\nu_{1}\nu_{2}}(x,e)  &  =(I_{e}^{2}F_{\mu_{1}\mu_{2}\nu_{1}\nu_{2}}%
^{s=2})(x)e^{\mu_{1}}e^{\mu_{2}}\\
A_{\nu_{1}\nu_{2}}^{(2)}(x,e)  &  =(I_{e}^{2}F_{\mu_{1}\mu_{2}\nu_{1}\nu_{2}%
}^{\left\vert h\right\vert =2})(x)e^{\mu_{1}}e^{\mu_{2}}%
\end{align}
This means in particular that the massive $s=2$ sl potential obtained by
fattening the $A^{(2)}~$is not the same as $A$ although both account correctly
for the $2s+1$ spin degrees of freedom and share their Wigner-Fock Hilbert
space. The massless limit of $A$ splits into the direct sum of the two
$\left\vert h\right\vert =2$ degrees of freedom and the $h=0$ contribution
which is the remnant of the $s_{3}=0$ component. Conserved currents and
stress-energy tensors preserve the number of degrees of freedom by converting
the $\pm s_{3}$ components into $\left\vert h\right\vert =s_{3}$ helicities of
a Wigner-Fock tensor product space.

In order to show how these results are related to the van Dam-Veltman-Zakharov
discontinuity problem one must look at some details. Whereas fattening and
taking the massless limit connect the 2-pfct of the 2-component massless
helicity $\left\vert h\right\vert =2$ potential$~A^{(2)}$ with that of its
5-component $s=2$ by deforming the momenta of the 2-pfct between
$H_{m}^{\uparrow}$ and $V^{\uparrow},$ the massless limit of $A$ is a cul de
sac from which a return to the original massive pure $s=2$ tensor potential is
not possible.

The relation between the massless limit of $A$ with that of $A^{(2)}$ are
easily seen to have the following form
\begin{align}
A_{\mu\nu}^{(2)}(x,e)  &  =A_{\mu\nu}(x,e)+\frac{1}{2}E_{\mu\nu}%
(e,e)A^{(0)}(x,e)\label{E}\\
E_{\mu\nu}(e,e)  &  =\eta_{\mu\nu}+(e_{\mu}\partial_{\nu}+e_{\nu}\partial
_{\mu})I_{e}+e^{2}\partial_{\mu}\partial_{\nu}I_{e}^{2}\nonumber
\end{align}
where the momentum space $E_{\mu\nu}~$has been rewritten as an
integro-differential operator acting on a scalar sl field and the massless
limit of $A^{(0)}~$is a (properly normalized) scalar escort. Combining this
relation with that between the $s=2~$pl field $A^{P}$, its sl counterpart $A$
and the derivatives of escorts (the $s=2$ analog of (\ref{escort})) one
obtains%
\[
A_{\mu\nu}^{P}=A_{\mu\nu}+\hbox{derivatives of escorts}
\]
one concludes that in the adiabatic limit the interaction between "massive
gravitons" and a trace-free energy-momentum tensor source $T_{\mu\nu}$ is
\cite{MSY2}
\begin{align}
\lim_{m\rightarrow0}\int A_{\mu\nu}^{P}T^{\mu\nu}  &  =\lim_{m\rightarrow
0}\int A_{\mu\nu}T^{\mu\nu}=\int(A_{\mu\nu}^{(2)}-\frac{\eta_{\mu\nu}}%
{\sqrt{6}}\varphi)T^{\mu\nu}\\
where\text{ ~}\varphi(x)  &  =\sqrt{\frac{3}{2}}\lim_{m\rightarrow0}%
a^{(0)}(x,e)
\end{align}
The independence of the integrated massless $A^{(2)}$ contribution from the
string direction follows from
\begin{align}
\partial_{e_{\kappa}}A_{\mu\nu}^{(2)}  &  =m^{-1}(\partial_{\mu}A_{\kappa\nu
}^{(2)}+\partial_{\nu}A_{\mu\kappa}^{(2)})\\
\partial_{e_{\kappa}}J_{\mu}^{~~\nu}  &  =-\frac{p_{\mu}}{(pe)_{i\varepsilon}%
}J_{\kappa}^{~~\nu}\nonumber
\end{align}
which in turn follows from the identity in the second line (for more details
see \cite{MRS2}) and represents the $s=2$ counterpart of the relation between
de Sitter space 1-forms in (\ref{c}).

The result confirms the van Dam-Veltman-Zakharov discontinuity: the massless
limit of massive gravity differs from the result obtained directly with
massless gravitons. But different from Zakharov's calculation which identifies
this contribution as being the relic of a unphysical gauge theoretical degrees
of freedom, the present calculation shows that it is really the massless
footprint of the physical $s_{3}=0$ spin component. For the traceless
stress-energy tensor of photons the last contribution vanishes whereas for
couplings to matter (mercury perihelion) it remains.

This calculation permits a straightforward extension to any spin. The relation
between the Proca potential, its sl counterpart and the associated sl escorts
reads
\begin{equation}
A_{\mu_{1}\dots\mu_{s}}^{P}=A_{\mu_{1}\dots\mu_{s}}+sym.(\partial_{\mu_{1}%
}\phi_{\mu_{2}\dots\mu_{s}}+\partial_{\mu_{1}}\partial_{\mu_{2}}\phi_{\mu
_{3}\dots}+\dots+\partial_{\mu_{1}}\dots\partial_{\mu_{s}}\phi) \label{sl}%
\end{equation}
where the $\phi_{\mu_{1}\dots\mu_{i}}~$is an $s-i$ fold iterated line integral
along $e$ of the spin $s~$Proca potential and the symmetrization is over all
indices and the $\phi$ are already symmetric by construction. For our purposes
it is more convenient to use a different basis of escorts which are obtained
by descending from the sl $A_{\mu_{1}\dots\mu_{s}}$ in terms of divergencies
\begin{align}
&  A_{\mu_{1}\dots\mu_{s}}^{P}=A_{\mu_{1}\dots\mu_{s}}-sym.(\frac
{\partial_{\mu_{1}}}{m}a_{\mu_{2}\dots\mu_{s}}^{(s-1)}+\frac{\partial_{\mu
_{1}}\partial_{\mu_{2}}}{m^{2}}a_{\mu_{3}\dots}^{(s-2)}+\dots+\frac
{\partial_{\mu_{1}}\dots\partial_{\mu_{s}}}{m^{s}}a^{(0)})\label{recur}\\
&  ma_{\mu_{r}\dots\mu_{s}}^{(s-r)}=-\partial^{\mu}a_{\mu\mu_{r}\dots\mu_{s}%
}^{(s-r+1)},~\ ~~a_{\mu_{1}\dots\mu_{s}}^{(s)}:=A_{\mu_{1}\dots\mu_{s}%
}\nonumber
\end{align}
The second line shows that the $a$ escorts start from the sl potential and
descend by differentiation instead of descending from $A^{P}$ by
line-integration. The $a$ have the same dimension $d_{sd}=1=d_{eng}%
,~d_{infr}=0$, and are linear combinations of the $\phi$ escorts. As long as
$m>0~$each escort carries the full content of the Wigner spin $s$~representation.

Although the $a^{\prime}s~$have a massless limit they still do not decouple.
The van Dam-Veltman-Zakharov discontinuity shows that for $s=2$ the
$\left\vert h\right\vert =2$ and $h=0$ contributions stay together and have to
be separated with the help of an integro-differential operation (\ref{E}). The
analogous situation in the general case is that the even and odd $s_{3}%
~$contributions remain coupled among themselves and can only be split in terms
of their helicity content by the use of such integro-differential operations
\cite{MRS2}. Naturally one can obtain a spin $s$ vector potential from
fattening a massless helicity $h$ potential if $h=s.$

The tensor $v_{\mu_{1}..\mu_{\left\vert h\right\vert }}(q,e)$ which appears in
the relation of the helicity $h$ tensor field $A_{\mu_{1}..\mu_{\left\vert
h\right\vert }}^{(\left\vert h\right\vert )}$ and the Wigner operator
$a^{\#}(q,h)$ (which extends the construction of $A_{\mu\nu}^{(2)}~$in
(\ref{E}) to arbitrary helicity $h).$ This $e$-dependent polarization tensor
$v_{..}(q,e)~$replace the only up to re-gauging defined polarization tensor.
If used in Weinberg's soft scattering limit of a massless particle with
momentum $q$ scattering on $n$ massive particles with momenta $p_{i},i=1,..n$
\cite[4.1]{Ra-Ta}, one obtains the same conclusions except that the gauge
theoretic argument is replaced by the $e$-independence which follows in first
order from the fact that the directional derivative with respect to $e$ on
these polarization tensors can be written as a spacetime derivative
$\partial_{\mu}$ acting on such a tensor in analogy to (\ref{6}). The use of
sl polarization tensors instead of gauge symmetry is required by using
positivity which guaranties the exclusive appearance of physical degrees of freedom.

The weakness of Lagrangian constructions of conserved currents and
stress-energy tensors is that with the exception of low spins there is no
guaranty that the so obtained classical expressions have the correct
commutation relations with the quantum fields. It is much safer and easier to
\textit{start from the commutation relations between Wigner's generators} of
the Poincar\'{e} group and the Wigner particle operators $a^{\#}(p,s_{3})$ and
to rewrite them with the help of the intertwiners into covariant commutation relations.

\subsection{Infinite spin revisited}

A simple illustration of such an "intrinsic quantum" construction of the
stress-energy tensor has been recently presented in \cite{PL}. One starts from
the expressions of the infinitesimal generators of translation $\mathbf{P}%
_{\mu}~$and Lorentz generators $\mathbf{M}_{\mu\nu}~$in terms of the Wigner
operators $a^{\#}(p,s_{3})$%
\begin{align}
\mathbf{P}_{\mu}  &  =\int\sum_{s_{3}}a^{\ast}(p,s_{3})p_{\mu}a(p,s_{3}%
)d\mu(p)~\label{P}\\
\mathbf{M}_{\mu\nu}  &  =-i\int(\delta_{s_{3}s_{3}^{\prime}}p\wedge
\partial_{p}+d(\omega)_{s_{3}s_{3}^{\prime}}^{t})_{\mu\nu}a^{\ast}%
(p.s_{3})a(p,s_{3}^{\prime})d\mu(p) \label{M}%
\end{align}
The first step is two rewrite the contribution of the spin component $s_{s}$
to$~\mathbf{P}_{\mu}\,$as
\begin{align}
&  \mathbf{P}_{\mu}=\int\int d\mu(p)d\mu(p^{\prime})\sum_{s_{3},s_{3}^{\prime
}}(p_{\mu}a^{\ast}(p,s_{3})\delta_{s_{3}s_{3}^{\prime}}(2\pi)^{3}%
\delta(\mathbf{p}-\mathbf{p}^{\prime})(p_{10}+p_{20})a(p^{\prime}%
,s_{3}^{\prime})\label{P1}\\
&  (2\pi)^{3}\delta(\mathbf{p}-\mathbf{p}^{\prime})=\int e^{-i(\mathbf{p}%
-\mathbf{p}^{\prime})x}d^{3}x=\int e^{-i(p-p^{\prime})x}d^{3}x \label{P2}%
\end{align}
where in the second line used the cancellation of the $p_{0}~$components.

What remains to do is to convert the Wigner operators via intertwiners into
the covariant fields. For this one uses their completeness relation in order
to write the unit operator in spin space as
\[
g^{MN}\nu_{Ms_{3}}\overline{\nu_{Ns_{s}^{\prime}}}=\delta_{s_{3}s_{3}^{\prime
}}%
\]
where $M$ and $N~$represent the multi-tensor indices of the intertwiner. What
remains is to use the Fourier transform (\ref{P2}) and pass from the Wigner
operators to the fields. Using the fact that the $a^{\ast}a^{\ast}\ $and $aa$
contributions vanish as a result of the presence of$~\overleftrightarrow
{\partial}_{0}$ and that $aa^{\ast}$ terms are absent in Wick-ordered products
one verifies that
\begin{equation}
\mathbf{P}_{\mu}=\int\tilde{T}_{\mu0}(x)d^{3}x,~~~\tilde{T}_{\mu\nu}%
(x)=-\frac{1}{4}\int:A_{\mu_{1}..\mu_{s}}^{P}(x)\overleftrightarrow{\partial
}_{\mu}\overleftrightarrow{\partial}_{0}A^{P,\mu_{1}..\mu_{s}}(x):
\end{equation}
where $\tilde{T}_{\mu\nu}$ is a contribution to the stress-energy tensor.

The full tensor density which generates all Poincar\'{e} transformations is of
the form%
\begin{equation}
T_{\mu\nu}=\tilde{T}_{\mu\nu}+\partial^{\rho}\Delta_{\mu\nu,\rho}%
\end{equation}
To compute the second contribution, which is also a bilinear expression in the
$A^{P}~$tensor fields, one starts from the bilinear expression for $M_{\mu\nu
}$ in terms of the $a^{\#}~$Wigner operators which also contains a
contribution the infinitesimal part of Wigner's little group. The
representation of the Poincar\'{e} group generators in terms of pl
stress-energy tensors may be rewritten in terms of their sl counterparts
\cite{MRS2}. For recent results about constructing infinite spin fields and
their E-M tensors as Pauli-Lubanski limits we refer to \cite{PL}.

Rehren's construction of infinite spin quantum fields in terms of the
Pauli-Lubanski limit is the most natural one; it corresponds to the use of the
distinguished tensor potentials obtained by fattening its unique massless
counterpart at fixed spin, except that it goes into the opposite direction at
fixed P-L parameter\footnote{In this way it selects a unique countable family
of fields within the equivalence class of all relatively causal fields
constructed in \cite{MSY2}.}. The tensor field disappears in this limit and
what remains (after appropriate adjustments) is the infinite family of escorts
with arbitrary high tensor degree.

The nonexistence of the infinite spin tensor potential $A_{\mu_{1}\mu
_{2}...\infty}$ accounts for the absence of a relation which converts the
differential $d_{e}$ into a spacetime divergence as well as the absence of a
gauge theoretic formulation. This is the reason why infinite spin matter
cannot interact with ordinary quantum matter \cite{dark}.

Of physical relevance is the existence of conserved currents and
energy-momentum tensor in the sense of bilinear forms \cite{PL}. Hence
expectation values of E-M tensors and possible gravitational backreaction
remain physically meaningful.

\section{Causality and the Velo-Zwanziger conundrum}

The Velo-Zwanziger conundrum is an alleged causality paradox which arose from
the naive expectation that $s\geq1$ quantum fields, whose free field equations
are modified by linear pl couplings to external potentials, maintain their
causal propagation. Formally it is closely related with the Weinberg-Witten
No-Go theorem which excludes the existence of higher helicity conserved pl
currents. This connection turns out to be useful for the solution of the V-Z conundrum.

\subsection{Recalling the solution of the Weinberg-Witten problem and the
associated local charges}

In \cite{MRS2} it was shown that for massive $s\geq$ $1$ free field one can
construct sl tensor potentials whose associated conserved sl currents have
finite massless limits even when according to the Weinberg-Witten (W-W)
theorem physical (gauge-invariant) pl currents do not exist.

In the massive case both the pl and sl currents are members of the same local
equivalence class which consist of all Wick-ordered composites of pl fields
and their related sl counterparts. Their relative causality reads
\begin{equation}
\left[  j_{\mu}^{P}(x),j_{\nu}(x^{\prime},e)\right]  =0\text{~}for\text{
}x\,{\lower1pt\hbox{\LARGE$\times$}}\,\mathcal{S}(x^{\prime},e),~\mathcal{S}%
(x^{\prime},e)=x^{\prime}+\mathbb{R}_{+}e,~e^{2}=-1\text{ }or~0 \label{ten}%
\end{equation}
Their charge-densities differ by spatial divergencies and hence they share the
global $U(1)~$generators. In the massless limit the sl spin potentials pass
continuously to their massless counterpart (not possible with pl potentials)
which act in the conformally covariant helicity Wigner-Fock space. The sl
currents are bilinear in the charge carrying sl potentials\cite{MRS2}.

The two currents (\ref{ten}) share the same "engineering" dimension (classical
dimension in terms of mass units) $d_{cl}=3,$ but possess different short
distance scaling dimensions $d_{sd}(j_{\mu}^{P})=2(s+1)+1$ and $d_{sd}(j_{\mu
})=d_{cl}=3;~$this accounts for the fact that the sl $j_{\mu}$ allows a
massless limit whereas $j_{\mu}^{P}$ diverges as $j_{\mu}^{P}\overset
{m\rightarrow0}{\sim}m^{-2s}\ $(the W-W obstruction). As expected, the sl
$j_{\mu}(x,e)$ admits a massless limit in which the $2s+1$ spin degrees of
freedom decompose into a direct sum of $s$ helicity and one scalar
contribution so that the Wigner-Fock space turns into a tensor product of
helicity spaces$.$

The presentation concerning the relation between pl and sl conservation laws
in \cite{MRS2} was mainly focussed on the stress-energy tensors (SET); in the
following we present the corresponding problem for conserved currents. A
convenient illustration is provided by the sl current with the lowest W-W
helicity $h=1$ as follows$.$\ 

Using the linear relation $A_{\mu}^{P}(x)=A_{\mu}(x,e)-\partial_{\mu}%
\phi(x,e)~$between pl and its canonically associated sl field and the gradient
of its escort derived in the previous section (\ref{escort}) one finds that
the pl and sl currents are related as (omitting Wick-ordering)%
\begin{align}
&  j_{\mu}^{P}=iA^{P\nu}(x)^{\ast}\overleftrightarrow{\partial_{\mu}}A_{\nu
}^{P}(x)=j_{\mu}(A(x,e))+j_{\mu}(a(x,e))+\partial^{\kappa}C_{\kappa\mu
}\label{j}\\
&  a(x,e)=m\phi(x,e),~C_{\kappa\mu}=iA_{\kappa}^{\ast}\overleftrightarrow
{\partial}_{\mu}\phi~+i\phi^{\ast}\overleftrightarrow{\partial}_{\mu}%
\partial_{\kappa}\phi+h.c.\nonumber
\end{align}
The first two contributions are conserved sl currents whose massless limit
correspond to the current of the complex $s=1~$sl field $A_{\nu}(x,e)$ (which
replaces the nonexistent pl W-W current), and that of a complex scalar field
$a(x)=\lim_{m\rightarrow0}a(x,e).$ The $m^{-2}$ W-W obstruction $C$ does not
contribute to the global charge.

The "obstructing" contribution $\partial^{\kappa}C_{\kappa\mu}$ carries both
the leading short distance dimension $d_{sd}=5~$and and the $m^{-2}$
divergence which is the culprit for the W-W problem. This kind of
decomposition into $s~$conserved$~d_{sd}(j)=3~$sl currents, which pass for
$m\rightarrow0$ to $s$ sl helicity currents and a pl current of a scalar
particle, exists for every spin $s\geq1.$

Using the free field equation for $A_{\mu}$ and $\phi$ one verifies that
$C$-contribution is of the form of a spatial divergence and hence does not
contribute to the infinite volume limit of the charges \cite{MRS2}%

\begin{equation}
Q(A^{P})=Q(A)+Q(a)
\end{equation}
i.e. the massless limit decomposes the three spin degrees into the $\pm1$
helicities of $A_{\mu}$ and $h=0$ carried by $a.~$Before this limit both sl
fields $A_{\mu}$ and $a$ account for the three $s=1$ degrees freedm.

For pl currents there exists extensive literature on the problem of relation
between conserved currents, local charges, and their global limits
\cite{K-R-S}\cite{E-S}\cite{S-St}\cite{Req}. The basic idea is to start from a
conserved current and define%

\begin{align}
Q  &  =\lim_{R\rightarrow\infty}Q(f_{R},f_{d}),~~Q(f_{R},f_{d}):=j_{0}%
(f_{R},f_{d})\label{test}\\
f_{R}(\mathbf{x})  &  =\left\{
\begin{array}
[c]{c}%
1\ \ \ \ \ \ \ ~\left\vert \mathbf{x}\right\vert <R\\
\ \ 0~~\ \ \left\vert \mathbf{x}\right\vert >R+r
\end{array}
\right.  ~\label{test2}\\
f_{d}(x_{0})  &  \geq0.~~suppf_{d}\subset\left\vert t\right\vert <d,~\int
f_{d}(x_{0})dx_{0}=1
\end{align}
One then uses the conservation law of the current to show that the commutator
$\left[  Q(f_{R},f_{d}),A\right]  $ for $A$ $\mathcal{\in A(O})~$is
independent of the choice of the smearing function $g(x)=f_{R}(\mathbf{x}%
)f_{d}(x_{0})$ as long as $\mathcal{O}$ remains inside their timelike extended
shell structure (\ref{test2}).

The local charge $Q(g)$ which measures the charge of an operator $A$ localized
in $\left\vert \mathbf{x}\right\vert <R$ converges towards the generator
$Q~$of the global $U(1)$ symmetry. The concept of a local charge content
becomes problematic in case of sl currents since the use of a rigid spacelike
direction $e$~does not allow the causal separation of $j_{0}(f_{R},f_{d})$
from the localization region of $A.$ The heuristic idea for achieving such a
separation would be to "comb" the strings emanating from the shell between $R$
and $R+r$ into different directions so that they remain causally separated
from $A$ $\mathcal{\in A(O}).$ But then the strings emanating from poins
inside the shell would have to move to spacelike infinity outside the larger
sphere and violate the localization \textit{inside} the larger sphere

In view of a recent proof of the so-called \textit{split property}
\cite{L-M-P-R}, which is known to secure the local implementation of global
symmetries in massless $h\geq1$ \cite{D-L}, there is no problem with the
existence of local charges for QFT's with global symmetries; what is not clear
is whether such charges can be described in terms of conserved currents.

Meanwhile K.-H. Rehren informed me that his student M. Heep constructed local
charges from sl currents by appropriate use of conformal transformations. The
idea is to construct a local charge operator localized in a half-space, that
is then mapped to a sphere by a conformal transformation. In this way the
strings become "curled" and end in the north pole.

Hence the W-W No-Go theorem excludes pl currents, but does not affect the
causal localizability of charges in arbitrary small spacetime regions.

\subsection{The V-Z conundrum arises from an incorrect implementation of
causality}

A simple class of models for a critical examination of the V-Z conundrum is
provided by linear couplings of conserved currents to external vector
potentials the relevant property of the sl current is its lowered short
distance dimension. A suitable setting for such problems is obtaind in terms
of Bogoliubov's definition of the S-matrix and interacting local fields in
terms of adiabatic limits of the Bogoliubov $S(g)$-functional\footnote{Our use
of the Bogoliubov's formalism is close to that in \cite{Du-Fr}\cite{Wre}%
\cite{F-R}}%

\begin{align}
~  &  S:=\lim_{g(x)\rightarrow g}S(g),~~~S(g)=T\exp i%
{\displaystyle\int}
g(x)L_{int}(x)d^{4}x\label{B}\\
&  A(x)|_{L_{int}}=\lim_{g(x)\rightarrow g}\frac{\delta}{i\delta f(x)}%
S^{-1}(gL)S(g(x)L_{int}(x)+fA)|_{f=0} \label{C}%
\end{align}

Here the interaction density $L_{int}$ is a Wick-ordered product of not more
than 4 free fields from the class of Wick-ordered composites of free fields
and $A|_{L_{int}}$ the interacting counterpart of $A(x)~$which is either a
free field or a Wick-ordered product of free fields (the terminology "free" is
used for linear fields and Wick polynomials). The interacting field has the
form of a power series in $g$ with retarded products of $n$ $L^{\prime}s$
which is retarded in $A(x).$ The linear Bogoliubov map $A\rightarrow
A|_{L_{int}}$ does not preserve the algebraic structure but it maintains the
property of causal separability. Hence fields constructed in this way maintain
causality, and the solution of the V-Z problem consists in the proper
computation of the interacting fields via (44).

The class of interactions with external potentials to be studied is of the
form $L_{int}=L^{P}=U^{\mu}j_{\mu}^{P}$ with $U^{\mu}$ a external (classical)
vector potential and $j_{\mu}^{P}$ a conserved current as before. For the
current of a scalar complex free field $\varphi$ there is no problem; its
conserved current has $d_{sd}(j_{\mu}^{P})=3$ and hence (with $d_{sd}%
(L_{int})=3$) renormalizable. This is the model on which Velo and Zwanziger
base their propagation picture: namely the scalar field obeys a linear field
equation which is linear\footnote{Here and in the sequel linear stands for
linear in $U_{\mu}$ and its derivtives.} in $U^{\mu}$ and they expect
(erroneously, as will be seem) that this holds independent of spin.

For $s=1$ the $d_{sd}(L^{P})=5$ and hence the pl model is nonrenormalizable.
To reduce the $d_{sd}$ from 5 to 4 one uses the relation (\ref{j}) which
rewritten in terms of the interaction density reads%

\begin{align}
~\text{ }  &  L^{P}=j_{\mu}^{P}U^{\mu}=L-\partial^{\kappa}V_{\kappa
},~C_{\kappa\mu}=iA_{\kappa}^{\ast}\overleftrightarrow{\partial}_{\mu}%
\phi~+i\phi^{\ast}\overleftrightarrow{\partial}_{\mu}\partial_{\kappa}%
\phi+h.c\label{m}\\
&  L:=j_{\mu}^{s}U^{\mu}-C_{\kappa\mu}\partial^{\kappa}U^{\mu},~~V_{\kappa
}=-C_{\kappa\mu}U^{\mu},\text{ }S^{(1)}\overset{a.l.}{=}\int L^{P}d^{4}x=\int
Ld^{4}x\nonumber
\end{align}
where the decomposition (\ref{j}) of $j_{\mu}^{P}$ was used. Since
$d_{sd}(C_{\kappa\mu})=4$ the power counting bound $d_{d}(L)=4$ holds, the
model is renormalizable and its first order S-matrix $S^{(1)}$ (the adiabatic
limit of the interaction density) is the same for the two densities and hence
string-independent (the suitably defined adiabatic limit of the $\partial V$ vanishes).

The decomposition of $j_{\mu}^{P}$ (\ref{j}), which previously served to solve
the W-W problem (by converting the pl current into its for $m\rightarrow0$
regular sl counterparts and a $C$-term, which carries the $m^{-2}$ mass
divergence but does not contribute to the global charge\footnote{Using
conformal invariance of massless helicity representations one can also show
the existence of local charges (see remarks in previous subsection).}), is now
used to solve the V-Z causality problem. To achieve this one uses the fact
that the $C$-term is a 4-divergence and disappears in the adiabatic limit
which represents the S-matrix.

The two interaction densities $L^{P}$ and $L$ share the same S-matrix; whereas
the pl $L^{P}(x)$ side insures that the S-matrix is that of a causal
interaction, the sl $L(x,e)$ guaranties the renormalizability of $S^{(1)}.$
The $L(x,e)~$together with $V_{\mu}(x,e)$ forms what will be referred to as a
$L,V_{\mu}$ pair. The first order S-matrix (\ref{m} second line), which is the
adiabatic limit of the interaction density, is the same for $L^{P}$ and $L~$
referred to as the linear relation (\ref{m}). The $L^{P}$ represents the
heuristic content of the interaction, but as a result of its bad short
distance behavior it is not suitable for perturbative calculations. The short
distance improved $L$ weakens the localization but retains enough of it to
keep fields causally separated and to maintain scattering theory.

The remaining problem is the extension of this idea to higher order. For
convenience of notation one uses a differential formulation of pl localization
in the form of $e$-independence in the form $d_{e}(L-\partial V)=0.$ It is
convenient to use lightlike $e^{\prime}s$ since in this case no smearing is
needed. The problem in higher order is the time-ordering. For the
$e$-independence of the S matrix one needs the $\partial$ to act outside the
time-ordering e.g.%
\begin{align}
&  d_{e}(TL(1)L(2)-\partial_{1}^{\mu}TV_{\mu}(1)L(2)-\partial_{2}^{\nu
}TL(1)V_{\nu}(2)+\partial_{1}^{\mu}\partial_{2}^{\nu}TV_{\mu}(1)V_{\nu
}(2)\overset{?}{=}0\label{indep}\\
&  ~d_{e}T(L(1)-\partial V(1))(L(2)-\partial V(2))=0\nonumber
\end{align}
and higher order extensions involving one $V_{\mu}$ and $n-1$ $L^{\prime}s.$

This is generally not possible without creating "obstructions" of the form of
delta contributions of the form $\delta(x_{1}-x_{2})d_{e}L_{2}(x_{1},e)~$which
are quadratic in $U_{\mu}$ and its derivatives. Higher order violations may
lead to contributions of higher polynomial degree in $U_{\mu}$ and
derivatives; it is a characteristic property of obstructions in models of
external potential interactions that all obstructions remain bilinear in the
quantum fields.

These obstructions are absorbed in the form of induced contributions into a
modified Bogoliubov formalism by defining%
\begin{equation}
L_{tot}=L+\frac{1}{2}L_{2}+..\frac{1}{n!}L_{n}+.. \label{obst}%
\end{equation}
where $L_{n}$ is of polynomial degree $n$ in $U_{\mu}$ and its derivatives and
remains quadratic in free fields. Note that induced contributions do not
increase the number of parameters and hence must be distinguished from
counterterms of pl renormalization theory.

In the $s=1$ model (\ref{m})~the $L_{2}$ contribution \textit{can
alternatively be encoded into a redefinition of time ordering}
\begin{align}
&  T_{0}\partial_{\mu}\phi(x_{1})\partial_{\nu}\phi(x_{2})\equiv\partial_{\mu
}\partial_{\nu}T_{0}\phi(x_{1})\phi(x_{2})\label{2pt}\\
&  T\partial_{\mu}\phi(x_{1})\partial_{\nu}\phi(x_{2})=T_{0}\partial_{\mu}%
\phi(x_{1})\partial_{\nu}\phi(x_{2})+icg_{\mu\nu}\delta(x_{1}-x_{2}%
)\nonumber\\
&  \partial^{\mu}T\partial_{\mu}\phi(x_{1})\partial_{\nu}\phi(x_{2}%
)-T\partial^{\mu}\partial_{\mu}\phi(x_{1})\partial_{\nu}\phi(x_{2}%
)=(1+c)\partial_{\nu}\delta(x_{1}-x_{2})\nonumber
\end{align}
and the validity of (\ref{indep}) for the $T\,\ $time-ordering requires to set
$c=-1$.~For $s>1$ The kinematic $T_{0}$ time-ordering contains more
derivatives and one has accordingly more $c^{\prime}s$ which must be
numerically adjusted in such a way that the $T$ time-ordering satisfies the
higher order pair requirements beyond (\ref{indep}).

The following side-remark maybe helpful for the later extension of SLFT to a
full QFT. The second order $A_{\mu}A^{\mu}\varphi^{\ast}\varphi$ term of
scalar QED within the new SLFT can either be viewed as induced or encoded into
a change of time-ordering for the derivatives of the complex scalar field. But
not all obstructions can be absorbed in this way. The $H$-selfinteractions of
the Higgs model is a \textit{genuine} second order induced term which results
exclusively from the implementation of the positivity and causality principle
of QFT (rather than from an imposed Mexican hat interaction).

The verification of the higher order pair relations will be deferred to a more
complete treatment of external potential problems. The expected result is:

\textbf{Conjecture }Couplings of conserved currents to external potentials
fulfill the higher order time-ordered $L,V_{\mu}~$relation. For $s=1$ the
resulting field equations are quadratic in $U_{\mu}$ whereas for $s>1$ they
are of infinite order (expected since $d_{sd}(L)\geq5$ ).$\ $

The form of the linear causal field equations (in particular the higher order
$U$ contributions) is determined by the form of the induced contributions.

The external potential formalism and its formal connection with the solution
of the W-W problem works in an analogous way for $s=2.$ The sl potential
$A_{\mu\nu}(x,e)$ has two escorts, a vector $a_{\mu}$ and a scalar escort $a$
which can be chosen in such a way that the operator dimension for all fields
is identical to their classical dimension $d_{cl}=1$ (or $3/2$ for
half-integer spin)%

\begin{align}
&  A_{\mu\nu}^{P}=A_{\mu\nu}+m^{-1}(\partial_{\mu}a_{\nu}+\partial_{\nu}%
a_{\mu})+m^{-2}\partial_{\mu}\partial_{\nu}a\label{dec}\\
&  j_{\mu}^{P}(x)=iA_{\kappa\lambda}^{P\ast}\overleftrightarrow{\partial_{\mu
}}A^{P\kappa\lambda}=j_{\mu}^{s}(x,e)+\partial^{\kappa}C_{\kappa\mu
}\nonumber\\
&  j_{\mu}^{s}(x,e)=iA_{\kappa\lambda}^{\ast}\overleftrightarrow{\partial
_{\mu}}A^{\kappa\lambda}-2ia_{\kappa}^{\ast}\overleftrightarrow{\partial_{\mu
}}a^{\kappa}+ia^{\ast}\overleftrightarrow{\partial_{\mu}}a\nonumber
\end{align}

In this case $d_{sd}(L^{P})=7$ and $d_{sd}(j^{\mu})=3~$and hence $L^{P}$ is by
$3$ units beyond the power-counting bound $d_{sd}=4$. For $s=1$~the
$\partial^{\kappa}C_{\kappa\mu}$ carries the highest $d_{sd}=7$
contribution.$~$After an additional linear disentanglement between $A_{\mu\nu
}$ and $a$ one arrives at a decomposition of $j_{\mu}^{s}$ which in the
massless limit represents the $h=2,1,0$ helicity contributions \cite{MRS2}.

The use of this decomposition in the rewriting of $L^{P}=j_{\mu}^{P}(x)U^{\mu
}$ for $s=2~$as a sl pair with $L^{P}=L-\partial V$ leads to a $d_{sd}%
(C_{\kappa\mu})=6~$ contribution which contains bilinear in $\phi=m^{-2}a$
terms with more than 2 derivatives. In analogy to counterterms in every order
in a nonrenormalizable full pl QFTs one expects to find induced terms in
arbitrary high orders.

It is worthwile to mention that there is also a gauge theoretic formulation in
which the linear operator relation between the sl potential and its pl
counterpart is replaced by the relation $A_{\mu}^{K}(x)=A_{\mu}^{P}%
(x)+\partial_{\mu}\phi^{K}$ where the $K$ refers to the Krein space and the
esort $\phi^{K}$ is the St\"{u}ckelberg negative metric pl scalar which adds
additional unphysical degrees of freedom to the indefinite Krein space. This
is the formulation of the Uni Z\"{u}rich group \cite{Scharf}\cite{Aste}
adapted to the presentation used in the present paper.

The gauge theoretic analog to the pair relation is $L^{K}=L^{P}+\partial^{\mu
}V_{\mu}^{K}.$ The model has a formal similarity with SLFT, but its pl
interpolating fields are unphysical; positivity obeying interpolating fields
are simply inconsistent with pl localization. The pair property works the same
way, one only has to replace the $d_{e}$ in (\ref{indep}) by the BRST
$\mathfrak{s}$.

However negative metric degrees of freedom lead to an unphysical realization
of appropriately nonlinear modified causal pl V-Z equation and should be
rejected inasmuch as gauge dependent pl currents have been discarted by W-W in
their No-Go theorem. Classical field theory is free of positivity requirements
and gauge theoretic causal propagation is perfectly compatible with its
principles. But the nonlinear dependence on external potential which was
overlooked by V-Z is also needed for the classical propagation of Cauchy date.

Recently there have been attempts to solve the V-Z conundrum in terms of
String Theory \cite{Por} \cite{Ra-Ta}. These authors extract a system of pl
equation in $d=3+1~$via dimensional reduction from the Virasoro algebra in 10
dimensions and found nonlinear modifications in case of constant external fields.

But the fact that there is nothing stringy about their pl equations raises the
old question: what do string-theorist really mean when they claim that their
objects are stringy in spacetime. Does their use of the terminology "string"
perhaps refer to a circular structur in a 10-dimensional target space whose
Fourier components correpond to the irreducible Wigner components of the
highly reducible superstring representation?

Their strings bear no relation with causal localization in spaceteime but
rather seem to refer to Born's quantum mechanical localization related to the
spectral decomposition of the \textbf{x} operator arises. Their use of
world-sheet and Nambu-Goto actions point into this direction and the way in
which they think of their localized objects as vibrating in space strengthens
this presumption. Causal localization in spacetime is very different (for more
see next section).

The next section explores important aspects of causal localization which,
although known to some experts, remained outside the conceptual radar screen
of most particle physicists.

\section{Particle wave functions and causal localization}

\textit{There is no concept in particle physics which led to more
misunderstandings than that of causal localization in spacetime. The strings
of String Theory obtained e.g. from quantized world-sheet or Nambu-Goto
actions bear no relation causal localization. A concept which reveals such
misunderstandings and corrects them in the clearest possible way is "modular
localization".}

\subsection{Newton-Wigner localization and its causality-providing modular
counterpart}

Wigner's theory of positive energy representations presents an interesting
meeting ground of two very different localization concepts. On the one hand
there is the quantum mechanical localization of dissipating wave packets whose
center moves on relativistic particle trajectories. Its formulation in terms
of quantum mechanical Born probabilities leads to the so-called Newton-Wigner
localization \cite{N-W}. For a scalar $m>0$ particle
\begin{align}
(\psi,\psi^{\prime})=\int\bar{\psi}\overleftrightarrow{\partial_{0}}%
\psi^{\prime}d^{3}x=\int\bar{\psi}_{NW}\psi_{NW}^{\prime}d^{3}x  & \\
\hbox{hence}\quad\tilde{\psi}_{NW}(\mathbf{p})=(2p_{0})^{-1/2}\tilde{\psi
}(\mathbf{p})  & \nonumber
\end{align}
Hence an improper N-W eigenstate of the position operator $\mathbf{x}_{NW}$
has a mass-dependent extension of the order of a Compton wave length. In
scattering theory, where only the large-time asymptotic behavior matters, such
ambiguities in assigning relativistic quantum mechanical positions at finite
times are irrelevant; the centers of wave packets of particles move on
relativistic velocity lines and the probability to find a particle dissipates
as $t^{-3}$ along these lines for all inertial observers. In fact Wigner never
thought of his Poincar\'{e} group representation theory as an entrance into
causal QFT; for him it remained part of relativistic QM\footnote{This perhaps
explains why Wigner, inspite of his overpowering role in the development of
$20^{th}$ century quantum theory, never participated in the QED revolution and
its QFT aftermath. For him his representation theory always remained part of
relativistic quantum mechanics (the quantum mechanical Newton-Wigner
localization in section 3). An interesting discussion can be found in Haag's
memoirs \cite{mem} page 276.}.

The more recent discovery of \textit{modular localization} shows that
causality properties are dormant within Wigner's positive energy
representation theory; they are reflected in properties of dense subspaces
obtained by applying algebras of local observables $\mathcal{A(O}$ to the
vacuum state $\mathcal{H(O})=\mathcal{A(O)}\Omega~$and projecting the so
obtained dense set of states of a QFT to the one-particle subspace
$\mathcal{H}_{Wig}(\mathcal{O})=E_{1}\mathcal{H(O}).$ That such spaces are
dense in the Hilbert space (and consequently their projection in the
one-particle subspace) is a special case of a surprising discovery made in the
early in the early 60s (the Reeh-Schlieder theorem \cite{St-Wi} \cite{Haag})
which showed that the omnipresence of vacuum polarization confers to QFT
\textit{a very different notion of localization} from that of Born's quantum
mechanical setting based on position operators.

The projection $\mathcal{H}_{Wig}\mathcal{(O)}~$has the remarkable property
that it can be constructed without the assistance of QFT \textit{solely in
terms of data from Wigner's representation theory} and that in the absence of
interaction one can even revert the direction and obtain the net of causally
localized subalgebras directly from that of modular localized Wigner subspaces
\cite{BGL}.

In this way one does not only gain a more profound understanding of QFT but
one also learns that Weinberg's pure group theoretic construction of
\textit{intertwiners} starting from Wigner's representation theory is part of
a much more general setting which, if properly used, leads to an extension of
perturbative renormalizability. This important concept of modular localization
was not available during Wigner's lifetime (see remarks in the introduction).

The simplest way to see that the quantization of a relativistic classical
particle associated with the action $\sqrt{-ds^{2}}$ does not lead to a
covariant quantum theory is to remind oneself that there exists no operator
$\mathbf{x}$ which is the spatial component of a covariant 4-vector
\cite{MSY2}. The conceptual problem one is facing is better understood by
first showing that causal localization bears no relation to Born's
probabilistic quantum mechanical definition.

Starting from the quantum mechanical projectors $E(R)$ for $R\subset
\mathbb{R}^{3}$ which appear in the spectral decomposition of $\mathbf{x}%
_{op}$
\begin{equation}
\mathbf{x}_{op}=\int\mathbf{x}dE(\mathbf{x}) \label{Ma}%
\end{equation}
one has
\[
E(R)E(R+\mathbf{a})=0\quad\hbox{for}\quad R\cap(R+\mathbf{a)}=0
\]
Define $E(R+a)=U(a)E(R)U(a)^{\ast}$ for $a\in\mathbb{R}^{4}.$ Assuming that
this orthogonality relation has a causal extension in the sense that $~$
$E(R)E(R+a)=0$ for spacelike separated$~R{\lower1pt\hbox{\LARGE$\times$}}%
R+a~$leads immediately to clash with positivity of the energy. This follows
from the fact that the positivity of the energy leads to the analyticity of
expectation values $(\psi,E(R)E(R+a)\psi)$ for $\operatorname{Im}a_{0}>0$
which in turn implies their identical vanishing (the Schwarz reflection
principle) and with$~\left\vert \left\vert E(R)\psi\right\vert \right\vert
^{2}=0$ the triviality of such projectors $E=0.$

A slight extension of the argument reveals that it can be dissociated from the
position operator of quantum mechanics. It then states that in models with
energy positivity\textit{ it is not possible to describe causal localization
("micro-causality") in terms of projectors and orthogonality of subspaces
}\cite{Malament}. A profound intrinsic understanding of causal localization in
QFT points into a very different direction from that obtained from the
quantization of actions describing classical world lines or world sheets and
Nambu-Goto action. The problems become insurmountable of one tries to
construct actions in the presence of several of such objects and their distance.

Before presenting the relation to QFT it is worthwhile to mention a little
known fact: it is perfectly possible to construct a relativistic description
of interacting particles in relativistic QM build on macro-causality. For two
particles this amounts to Poincar\'{e} group preserving modification in the
centre of mass system, but for more particles it is more complicated
(\cite{SHPMP} section 3). Apart from the fact that it leads to a
Poincar\'{e}-invariant S-matrix, it does (unlike Schr\"{o}dinger theory) not
permit a description in terms of second quantization.

\subsection{Mathematical properties of modular localization}

To prepare the ground for causal localization it is helpful to start with some
mathematical concepts concerning relations between real subspaces $H$ (linear
combination with reals) of a complex Hilbert space $H\subset\mathcal{H}%
\mathbf{.}$ The symplectic complement $H^{\prime}~$of a real space is defined
as the closed real subspace ($\overline{H^{\prime}}=H^{\prime}$) defined in
terms of the imaginary part of the scalar product in$~\mathcal{H}$%
\begin{align}
H^{\prime}  &  =\left\{  \xi\in\mathcal{H};\ \operatorname{Im}(\eta
,\xi)=0~\forall\eta\in H\right\} \label{sym}\\
H_{1}  &  \subset H_{2}\text{ }\Rightarrow H_{1}^{\prime}\supset H_{2}%
^{\prime}%
\end{align}
which turns out to be the real orthogonal space on the real $iH$ (only real
linear combinations)

A closed real subspace $H$ is called "standard" if it is both cyclic and
separating%
\begin{align}
H\quad\hbox{cyclic:}\quad &  \overline{H+iH}=\mathcal{H}\label{stan}\\
H\quad\hbox{separating:}\quad &  H^{\prime}\cap H=\left\{  0\right\}
\nonumber\\
\left(  H+iH\right)  ^{\prime}  &  =H^{\prime}\cap iH^{\prime}\nonumber
\end{align}
Cyclicity and the separation property have a dual relation in terms of
symplectic complements as written in the third line.

It is quite easy to obtain such standard spaces from covariant free fields. In
the simplest case of a scalar field the Hilbert space $\mathcal{H}$ is the
closure of the 1-particle Wigner space defined by the two-point function of
the smeared fields%
\begin{align}
\left(  f,g\right)   &  =\left\langle A(f)^{\ast}A(g)\right\rangle =\int
\tilde{f}^{\ast}(p)\tilde{g}(p)d\mu(p)\label{com}\\
&  \left[  A(f)^{\ast},A(g)\right]  =-i\operatorname{Im}\left(  f,g\right)
\nonumber
\end{align}
where $d\mu\ $is the invariant measure on the positive mass
hyperboloid.\ According to the Reeh-Schlieder theorem \cite{Haag} the
one-particle projection of the dense subspace of causally $\mathcal{O}%
$-localized states\footnote{States localized in the spacetime region
$\mathcal{O}$ are defined as the dense\ Reeh-Schlieder subspace$\ $%
obtained$\ \ $as $\mathcal{A(O)}\left\vert 0\right\rangle $ where
$\mathcal{A(O)}$ is the $\mathcal{O}$ localized subalgebra of $\mathcal{A}$.}
is dense in the one-particle Wigner space.$\mathcal{\ O}$-localized real
testfunctions define a dense real subspace $H(\mathcal{O)}$ and causal
disjointness corresponds to "symplectic orthogonality" and produces a
\textit{closed} real subspace.%
\begin{equation}
H(\mathcal{O}^{\prime})=H(\mathcal{O)}^{\prime}%
\end{equation}

As a side remark we mention that the same construction applied to a higher
halfinteger spin field leads to a corresponding situation
\begin{equation}
ZH(\mathcal{O}^{\prime})=H(\mathcal{O})^{\prime},\ \ Z=\frac{1+iU(2\pi)}{1+i}
\label{Z}%
\end{equation}
where the unitary "twist" operator $Z$ which is related to the factor $-1$ of
the $2\pi$ rotation. The use of the twist operator allows to treat bosons and
fermions under one common roof.

The important step for an intrinsic understanding of QFT (i.e. without the use
of P. Jordan's "(quasi-)classical crutches" of quantization) is to
\textit{invert the previous construction: find a relation within Wigner's
representation theory which permits to define }$\mathcal{O}$\textit{-localized
real subspaces which have the correct covariant transformation properties
under Poincar\'{e} transformations \cite{BGL}.} For this purpose it is helpful
to reformulate the above properties so that they take the form known from the
mathematical Tomita-Takesaki theory of operator algebras which permits a
direct connection of positive energy Wigner representations with a "local net
of operator algebras". It provides a unified view in which Weinberg's pl
intertwiner formalism and its sl extension are seen as two ways of generating
the same free field theory. As shown in previous sections the improvement of
short distance properties is the basis for a new perturbative renormalization setting.

To achieve this one needs an additional mathematical tool. The first step
consists in a suitable extension of the modular concepts. A standard subspace
$H$ comes with a distinguished operator. With $D(Op)$ denoting the domain of
definition of an operator one defines

\begin{definition}
A Tomita operator $\mathcal{S}$ is a closed antilinear densely defined
involutive operator $D(\mathcal{S})\subset\mathcal{H}$
\end{definition}

In physics one encounters such "transparent" operators only in QFT. It is easy
to see that there exists a 1-1 correspondence between Tomita operators and
standard subspaces $H;$ $H\leftrightarrow\mathcal{S}$. This follows from the
definition $\mathcal{S}(\xi+i\eta)=\xi-i\eta,~\xi,\eta\in H,$ whereas the
opposite direction is a consequence of the definition $H=\ker\left\{
\mathcal{S}-1\right\}  .$

As a result of involutiveness, the full content of Tomita operators is
contained in their dense domains which coincides with their range
("transparency"). Hence modular theory may be alternatively formulated in
terms of subspaces (\textit{\ref{stan}}) $Dom\mathcal{S}$ of Tomita operators.

The polar decomposition of $\mathcal{S}=J\Delta^{1/2}$ of $\mathcal{S}$ into
an anti-unitary $J$ and a positive operator $\Delta\ $with $D(S)=D(\Delta
^{1/2})$ leads to the unitary modular group $\Delta^{it}$ acting in
$\mathcal{H}$ and preserving the standard subspace $\Delta^{it}H=H,$ whereas
the modular conjugation maps into the symplectic complement $J$ $H=H^{\prime}$

A Tomita operator appears in a natural way in Wigner's representation theory
of positive energy representations of the Poincar\'{e} group $\mathcal{P}$. It
is obtained by defining $\Delta_{W_{0}}^{it}$ in terms of the Lorentz boost
operator which leaves the wedge $W_{0}=\left\{  x;~z>0,\left\vert t\right\vert
<z\right\}  $ invariant
\begin{align*}
&  \Delta_{W_{0}}^{it}=U(\Lambda_{W_{0}}(-2\pi t))\\
&  \mathcal{S}_{W_{0}}=J_{W_{0}}\Delta_{W_{0}}^{1/2},\ \ J_{W_{0}}=TCP\cdot
R_{\pi}%
\end{align*}
together with an anti-unitary $J$ obtained by multiplying the TCP reflection
$TCP$ with a $\pi$-rotation $R$ in the $x$-$y$\ plane as written in the second
line, taking $W_{0}~$into its causal complement.

The charge conjugation $C~$maps an irreducible Wigner particle space into its
charge conjugate and may need a doubling of the Wigner space whereas $TP$
corresponds to the spacetime inversion $x\rightarrow-x$. The preservation of
energy requires the time-reversal $T$ to be anti-unitary. Massless
representations need a helicity doubling $\pm h.$

Unbounded operators $\mathcal{S}$ whose dense domain is stable ("transparent"
in the sense \textit{domain=range}) are somewhat unusual in quantum physics;
they appeared first in quantum statistical mechanics \cite{Haag} and later in
searches for an intrinsic understanding of causal localization of QFT (without
referring to Lagrangian quantization) \cite{BW}. In the present context the
dense subspace of the Wigner space (possibly doubled by charge conjugation)
corresponds to wave functions which are "modular localized" in the wedge
$W_{0}$. Modular localization of Wigner wave functions is closely related to
causal localization of fields and provides an extension of the Weinberg
intertwiner formalism which includes Wigner's infinite spin class \cite{MSY2}.

\textit{The construction proceeds as follows: start from Wigner's positive
energy representation theory, define the Tomita operator }$S_{W_{0}}%
~$\textit{in the way described before, use the Poincar\'{e} transformations to
construct a net of modular localized real subspaces }$H(W)$\textit{ and use
the second quantization functor (the Weyl~or CAR functor) to pass to an
interaction-free net of standard wave functions spaces }$H(W)$ to
\textit{causally localized operator algebras }$\mathcal{A(}W)$ \textit{acting
in a Wigner-Fock Hilbert space.}

One also may directly construct real dense subspaces $H_{\mathcal{O}}$ and
their complexified counterparts $D(\mathcal{S}_{\mathcal{O}})=H_{\mathcal{O}%
}+iH_{\mathcal{O}}\subset\mathcal{H}$ corresponding to more general causally
complete convex spacetime regions $\mathcal{O}_{c}$ as intersections:%
\begin{align}
H_{\mathcal{O}_{c}}  &  =\bigcap_{W\supset\mathcal{O}_{c}}H_{W}\\
H_{\mathcal{O}}  &  =\bigcup_{\mathcal{O}_{c}\subset\mathcal{O}}%
H_{\mathcal{O}_{c}}\nonumber
\end{align}
whereas for more general regions the standard space is defined in terms of
exhaustion from the inside (second line). For details we refer to \cite{BGL}.

The energy-positivity of the massive and the $\pm\left\vert h\right\vert $
massless Wigner representation classes plays an important role in establishing
the isotony and causal localization of the "net of modular localized standard
spaces"%
\begin{align}
\quad\hbox{isotony:}\quad &  H_{\mathcal{O}_{1}}\subset H_{\mathcal{O}_{2}%
}\quad\hbox{if}\quad\mathcal{O}_{1}\subset\mathcal{O}_{2}\\
\quad\hbox{causality:}\quad &  H_{\mathcal{O}_{1}}\subset H_{\mathcal{O}_{2}%
}^{\prime}\quad\hbox{if}\quad\mathcal{O}_{1}{\lower1pt\hbox{\LARGE$\times$}}%
\mathcal{O}_{2}\nonumber
\end{align}
where ${\lower1pt\hbox{\LARGE$\times$}}$ denotes spacelike separation.

In the absence of interactions the passage from the spatial modular theory to
its algebraic counterpart is almost trivial. One passes to the Wigner Fock
space created by symmetrized tensor products and defines the $\mathcal{O}%
$-localized operator algebra as the von Neumann algebra generated by the Weyl
operators
\begin{align}
&  \mathcal{A(O})=\left\{  e^{iA(h)};h\in H_{\mathcal{O}}\right\}
^{\prime\prime},~\left[  A(h_{1}),A(h_{2})\right]  =i\operatorname{Im}%
(h_{1},h_{2})\label{Weyl}\\
&  \quad\hbox{with}\quad A(h)=\int\sum_{s_{3}=-s}^{s}(h(p,s_{3})a^{\ast
}(p,s_{3})+h.c.)d\mu(p),~h\in H_{\mathcal{O}}%
\end{align}
in terms of the Wigner creation/annihilation operators\footnote{The "second
quantization" counterparts of the Wigner wave functions.} $a^{\#}(p,s_{3})$ or
their helicity counterparts. For half-integer spin or helicity the presence of
the twist operator (\ref{Z}) leads to fermionic commutation relation. In both
cases the second quantization functor maps the spatial Tomita operator into
its algebraic counterpart which acts in the ("second quantized") Wigner Fock
space $\mathcal{H}$ associated with the Wigner one-particle space $H$
\[
\mathcal{S}_{\mathcal{O}}^{alg}A\Omega=A^{\ast}\Omega,~A\in\mathcal{A(O})
\]
In the sequel only the algebraic Tomita-Takesaki theory will be used and the
superscript $alg~$will be omitted for convenience.

Modular localized spaces $H(\mathcal{O})$ and their associated noninteracting
field algebras $\mathcal{A(O})$ are by construction "causally complete"
$\mathcal{A(O})=\mathcal{A(O}^{\prime\prime})$ and hence a fortiori Einstein
causal (section 2 \ref{Haag}). The causal completion $\mathcal{O}%
^{\prime\prime}$ is obtained by taking twice the causal complement
$\mathcal{O\rightarrow O}^{\prime}.$

For wedge regions the modular group coincides with the unitary Wigner
representation of the wedge-preserving Lorentz group; for all other regions
the modular groups in $m>0$ Wigner representations acts in a non-geometric
("fuzzy") way. It is believed that this nongeometric action$~$becomes
asymptotically (near the boundary of the causal completion) geometric.

Massless finite helicity theories have a larger set of regions related to
Huygen's principle in which modular groups act in a geometric way; this
includes all regions which are images of wedges under the action of conformal
transformation as e.g.\ finite double cones. Sl potentials whose semiinfinite
spacelike lines remain inside wedges may under suitable conformal
transformation pass to potentials localized on finite elliptic curves which
connect two points on different edges of double cones; they can be viewed as
substitutes for the nonexistent pl coordinatization of double cones.

It may also happen that the standard spaces for compact spacetime regions
$\mathcal{O}$ are trivial $H_{\mathcal{O}}=\left\{  0\right\}  .$This occurs
precisely for the Wigner's \textit{zero mass infinite spin representation} for
which the tightest localized nontrivial spaces correspond to modular
localization in arbitrary narrow spacelike cones. On the other hand zero mass
\textit{finite helicity} spaces are the most geometric representation since
their modular groups continue to act geometrically even for double cones; in
fact they correspond to conformal transformations which preserve double cones.
Unlike finite helicity fields \textit{infinite spin} representations are
massless but \textit{not conformal}.

Modular localization plays also an important role in the understanding of
topological peculiarities of massless $h\geq1~$free QFT in connection with
toroidal regions (thickened Wilson loops). Last not least without their use it
would not have been possible to discover the intrinsic noncompact localization
of Wigner's infinite spin matter and the string-like nature of its generating
causally localized fields.

\textit{This raises the question if modular theory preserves its constructive
power in the presence of interactions.} It turns out that in that case one is
required to use the stronger operator algebraic modular theory. Interacting
theories share the same modular groups $\Delta_{\mathcal{W}}^{it}$ with their
noninteracting counterpart which is solely determined by the particle content.

The interaction enters through the dependence of the $J_{\mathcal{W}}$ on the
interaction which, using the fact that the incoming TCP is related with its
outgoing counterpart through the S-matrix $S$~\cite{Jost} \cite{AOP}, amounts
to%
\begin{equation}
J_{\mathcal{W}}=J_{\mathcal{W}}^{(in)}S \label{S}%
\end{equation}
Again the starting point is a von Neumann operator algebra $\mathcal{A}$
acting in a Hilbert space $\mathcal{H}~$which contains a vector $\Omega$ which
is cyclic and separating under the action of $\mathcal{A}$ (in QFT \textit{the
Reeh-Schlieder property} for $\mathcal{A(O)}$)
\begin{align}
\quad\hbox{cyclic:}\quad &  \mathcal{A(O)}\Omega\quad\hbox{is dense in}\quad
\mathcal{H}\label{sub}\\
\quad\hbox{separating:}\quad &  A\Omega=0\quad\Rightarrow\quad A=0,~~A\in
\mathcal{A(O)}~\nonumber
\end{align}
The definition$~H$ $=\overline{\mathcal{A}_{sa}(\mathcal{O})\Omega}$ ($sa=$
selfadjoint part) or $\mathcal{H=}\overline{\mathcal{A(O)}\Omega}$ connects
the algebraic modular theory with its previously presented spatial
counterpart. But there remains an important difference: the map between
standard subspaces and algebras is generally not injective whereas the
algebraically generated subspaces (\ref{sub}) always are .

The interaction-free situation remains exceptional in that there exists a
functorial relation between modular localized Wigner subspaces and
interaction-free causally localized subalgebras defined in terms of the Weyl
map (\ref{Weyl}). This functorial map is lost in the presence of interactions
in which case the relation between modular theory and particles becomes more involved.

\subsection{A critical perspective based on modular localization}

Most of what is presently known about modular theory comes from the
Bisognano-Wichmann theorem \cite{BW} which clarifies the modular properties of
wedge-localized subalgebras $\mathcal{A}(W)$. In models with a complete
particle interpretation one can use the modular theory of free fields to
derive the B-W theorem in the presence of interactions \cite{JensBW}. Of
special interest is the relation of the modular conjugation $J$ with the TCP
operator which is known to be connected to the S-Matrix \cite{Jost}. This is
particularly useful in $d=1+1$ integrable models whose S-matrix is known
(\ref{S}).

For integrable models without bound states this led to an interpretation of
the Zamolodchikov-Faddeev algebra in terms of modular localization in which
the concept of "vacuum-polarization-free generators" (PFG)\footnote{The
weakest assumptions under which PFG's exist were determined in \cite{Mu}.}
plays an important role \cite{AOP}\cite{BBS}. This in turn led to existence
proofs for certain $d=1+1$ integrable models in terms of operator algebraic
constructions based on modular theory \cite{Lech}. For a recent account with
many references to previous publications see \cite{AL}. These results
complement those obtained in terms of the "bootstrap-formfactor" program in
\cite{Kar}.

The important role of the S-matrix for modular localization in wedges has
triggered attempts to reconstruct a full causal QFT from its "on-shell
footprint" in form of its S-matrix \cite{Bert1,Bert2}. These ideas are
presently too weak for constructions in higher dimensions. Among the
unexpected results of modular theory is the proof that models in $d=1+2$ with
anyonic statistics (or its nonabelian "plektonic" counterpart) have always
nontrivial S-matrices \cite{M-Br}.

Whereas the idea that a nonperturbative QFT is uniquely determined by its
S-matrix remains still part of folklore, the S-matrix based perturbative SLFT
construction in the previous section is the basis of the new perturbation
theory. Different from the standard approach based on Lagrangian quantizations
in which the S-matrix is obtained from the mass-shell restriction (the LSZ
reduction formula) of time-ordered correlation functions of fields, SLFT
inverts this situation by encoding the model-defining particle content into an
interaction density of a perturbative defined S-matrix. This part of the
construction uses exclusively pl or sl free fields which are directly related
to the particles. Interacting (off-shell) quantum fields are constructed in a
second step in which the higher order contributions to the interaction density
(which were induced in the construction of $S$) serve as an input (\ref{C}).

Each pl or sl field from the equivalence of free fields and their Wick-ordered
composites has an interacting counterpart. Interacting fields which have
nonvanishing matrixelements between the vacuum and a one-particle state have
after appropriate normalization the same large-time in- and out- limits.

Finally we come to an important point whose clarification was promised at the
beginning of this section namely the possible connection of sl quantum fields
with String Theory. String theorists attribute a string-like localization to
their objects without providing arguments in favor of causal localizability.
Ideas based on worldlines, worldsheets, Nambu-Goto actions or strings start
from classical relativistic mechanics and hope that quantization preserve
these properties. This is evident from the way in which the covariant
world-line action $\sqrt{-ds^{2}}$ in \cite{Po} is used to prepare the ground
for the subsequent presentation of world-sheets and Nambu-Goto actions. The
impossibility to place two of such vibrating quantum mechanical strings into a
relative spacelike position reveals the problems which ST has with causal
localization in spacetime.

Fact is that, apart from Lagrangian quantization of $s<1$ fields, only the
Wigner representation theory (which cannot be accessed by quantization)
contains the seed for causal localization. Modular localization as a pre-form
of causal localization needs positivity and is inconsistent with gauge theory.
One cannot declare an object as "stringy" because its classical action
suggests this. This type of misunderstanding is clearly visible in
Polchinski's use of the action of a relativistic particle as a preparatory
step for relativistic worldsheet and Nambu-Goto actions.

This does not exclude the use of the quantum mechanical Newton-Wigner
localization to describe the dissipation of wave-packets. One may even
construct a macro-causal Poincar\'{e}-invariant multi-particle theory which
satisfies cluster decomposition properties and leads to a Lorentz-invariant
scattering matrix (\cite{SHPMP} section3). Such a construction disproves the
conjecture that relativistic particles and cluster properties alone will lead
to QFT.

Proposals which avoid quantization of classical actions and try to find causal
quantum matter in the target space of certain models of $d=1+1~$conformal QFTs
are less easy to dismiss. One such proposal is based on the Virasoro-algebra
of a supersymmetric 10-component chiral conformal current model (the
"superstring"). Its target space contains an algebraic structure which leads
to a highly reducible unitary Wigner representation of the Poincar\'{e} group
\cite{Brow}.

This is primarily a group theoretical observation which (apart from the ten
spacetime dimensions) fits well into Majorana's 1932 project of finding
algebraic structure which, in analogy to the $O(4,2)$ hydrogen spectrum,
describes wave functions of families of higher spin/helicity particles. But
why should one believe that the corresponding fields are sl in the absence of
any supporting argument? Is the terminology perhaps related to picturing the
\textit{tower} of representations containing different masses and spins as
Fourier components of a kind of \textit{internal} circle? In that case the use
of "string" for something bears no relation to spacetime and would be misleading.

Looking at the ST literature one gets frustrated about the disproportionate
relation between its conceptual poverty as compared to its mathematical
richness which its vague physical pictures lead to in the hands of
mathematicians. The word "string" should be more than an "epitheton ornans"
for a physically insufficiently understood mathematical formalism.

With additional conceptual care one can also avoid a widespread
misunderstanding in the physical interpretation of the $AdS_{n+1}$-$CFT_{n}$
isomorphism. It was certainly consequential to complement the observation of
equality of the symmetry groups of the two spacetimes by the verification of a
stronger Einstein causality-preserving isomorphism between the two QFTs. But
unlike classical field theory the timelike completion property\footnote{The
causal completion $\mathcal{O}^{\prime\prime}$ is the result of taking twice
the causal complement.} in QFT $\mathcal{A(O}^{\prime\prime})=\mathcal{A(O}),$
(which roughly speaking describes causal propagation) \textit{is not a
consequence of the spacelike Einstein causality}. Formally it is equivalent to
Haag duality $\mathcal{A(O}^{\prime})=\mathcal{A(O})^{\prime}$ from which it
results by taking the commutant on both sides and rewriting $\mathcal{A(O}%
^{\prime})^{\prime}$ by applying Haag duality to the (generally noncompact)
region $\mathcal{O}^{\prime}.$

In the old days \cite{H-S}\cite{Landau} it was shown that Einstein causality
and the causal dependency property (formally equivalent to Haag duality) are
\textit{independent} requirements. Causality without Haag duality occurs if
there are "too many" degrees of freedom as in the case of the generalized free
field; a phenomenon which has no classical analog (since the notion of quantum
\textit{degrees of freedom} has no classical counterpart).

This manifests itself in a sort of "poltergeist effect" in that there may be
\textit{more quantum degrees of freedom streaming into the dependency region}
\textit{as time moves on} \textit{than there were in the original
appropriately defined initial ("Cauchy") data}; in operator-algebraic notation
$\mathcal{A(O})\varsubsetneq\mathcal{A(O}^{\prime\prime})$ while Einstein
causality$\mathcal{A(O})\subset\mathcal{A(O}^{\prime})^{\prime}$ is preserved.

In fact it is quite easy to construct Einstein-causal models for which this
completion property is violated. Generalized free fields with a suitably large
$\kappa$ behavior of their Kallen-Lehmann spectral functions (containing a
much larger cardinality of degrees of freedom than free fields) were used at
the beginning of the 60s\footnote{I am sometimes asked about the origin of
this terminology. The answer is simple, it accounts for the fact that Haag had
this idea already before I entered the collaboration; my contributions
consisted in providing calculations involving generalized free fields.} to
show that the \textit{causal shadow property} represents a separate
requirement (initially called "the time-slice property").

The heuristic picture is that "squeezing" a QFT with the natural cardinality
of degrees of freedom corresponding to a \textit{n+1}-dimensional QFT into an
\textit{n} dimensional spacetime ("holographic image") causes an
"overpopulation"; this is precisely what happens in the $AdS_{n+1}$-$CFT_{n}$
case \cite{D-R}. The simplest illustration is obtained by projecting a free
AdS field and noting that the resulting conformal field is a generalized free
field of the kind used in \cite{H-S}\cite{Landau}. In the \textit{opposite}
direction i.e. starting from a "normal" CFT one expects an "anemic" degree of
freedom situation on the AdS side. As shown in \cite{Rehren} this is precisely
what happens; in fact there are no degrees of freedom at all in compact AdS
regions (double cones); to find any one has to pass to infinitely extended
wedge-like regions in AdS.

The overpopulation of degrees of freedom distinguishes "holographic
projection" from "normal" lower dimensional QFT and this explains the
quotation marks in "pathology"; a holographic projection maintains the degrees
of freedom of the original QFT and in this way prevents the conformal side to
be a normal QFT. This is a point which had been ignored in most post Maldacena work.

A helpful viewpoint concerning such "overpopulated" models which one obtains
by dimension reducing projections is to not consider them as autonomous QFTs
but to view them rather as stereographic projections of the original QFT.

The degree of freedom issue is not limited to the AdS-CFT isomorphism but
\textit{affects all attempts to extend (quasi)classical Kaluza-Klein
dimensional reductions of lowering extra dimensions to full-fledged causal
QFT}. To the extend that such pictures are compatible with the principles of
QFT\footnote{\textit{Massaging} Lagrangians is not the same as passing from
one QFT to another.} they correspond to perform a stereographic projection on
the original theory and not to pass to a lower dimensional QFT. This poses the
question whether the fashion around extra dimensions would have occurred with
a higher conceptual awareness about the subtle nature of positivity and causal
localization of relativistic quantum matter which forbid the use of naive
quasiclassical arguments.

A breakdown of Haag duality for entirely different reasons occurs in local
nets generated by massless helicity $h\geq1~$free field strengths. This
"topological duality violation" was mentioned in section 2 where its relation
to the Aharonov-Bohm effect and linking numbers was explained. It is lost in
the positivity-violating gauge theoretical setting which cannot distinguish
between the Haag duality and the somewhat coarser Einstein causality.

In his book on Local Quantum Physics Haag proposes an interesting extension
from the established one-fold duality for observable algebras to causally
separated two double cone localized algebras. In fact he explores the
possibility of the existence of a homomorphism from the orthocomplemented
lattice of causally complete regions in Minkowski space into that of von
Neumann algebras of observables (\cite{Haag}, Tentative Postulate 4.2.1). For
a region which consists of two spacelike separated double cones $K_{1},K_{2}$
this requires $\mathcal{A}(K_{1}\vee K_{2})=\mathcal{A}(K_{1})\vee
\mathcal{A}(K_{2}).$

Haag notices with a certain amount of disappointment that duality is violated
for doubly localized observable algebras associated to the conserved currents
of free fields. This follows from the existence of a pair $\psi(f)\psi
(g)^{\ast}~$with $supp~f\subset K_{1},~supp~g\subset K_{2}~$which commutes
with $\mathcal{A}(K_{1}\vee K_{2})~$but is not in $\mathcal{A}(K_{1}%
)\vee\mathcal{A}(K_{2}).~$He viewed this as a shortcoming of free fields which
he expected to disappear in the presence of interactions. He uses the idea of
a "gauge bridge" between $\psi(f)$ and$~\psi(g)^{\ast}$ as a hint that a
future positivity-maintaining replacement of gauge theory may satisfy this
strengthened form of causality ("Haag duality") may.

The string-bridges of SLFT do precisely this i.e. they prevent that Haag
duality of two causally separated double cones is violated by a
charge-anticharge pair. In constrat to gauge bridges which have no material
content (since they can be changed by gauge transformation implemented by
indefinite metric gauge charges) string-bridges consist of quantum matter.
Hence the existence of string bridges provides a local method to distinguish
an interacting net from a free one.

Another somewhat more metaphoric way of saying the same is that SLFT results
from gauge theory by applying Occam's razor to indefinite metric- and ghost-
degrees of freedom. Gauge theory is its best placeholder within the setting of
Lagrangian quantization.

Interestingly the same string bridges which save Haag duality of observables
also allow to view interpolating fields as resulting from space- or light-like
limits in which the anticharge component is disposed of at infinity but leaves
a trail of quantum matter behind.

Whereas all sl fields in the absence of interactions can be obtained as
semiinfinite line integrals from pl fields \footnote{Note that the interaction
density and the S-matrix only uses such free fields.}, this breaks down in the
presence of interactions. In that case the necessarily sl localized
interpolating $s<1$ fields receive their sl localization metaphorically
speaking through being "infected" by their contact with higher order $s\geq
1~$sl potentials with which they share the interaction density. This will be
exemplified in a number of models in the next section.

For anybody who has grown up with Haag's way of looking at QFT it is deeply
satisfying that the model-dependent division between gauge invariant
observables and gauge dependent interpolating fields corresponds to the SLFT
localization dichotomy between pl observables and sl interpolating fields.

The fault line, which unfortunately still separates the ST community from
those who are working on the successful but still largely unfinished project
of QFT, runs precisely alongside the issue of causality and its refinements.
The recent progress on entanglement entropy \cite{Wi} \cite{Ho-Sa} requires a
profound understanding of causality in the context of operator algebras. The
fact that these ideas are presently also leading to a revision of perturbative
QFT raises hopes that this schism will be overcome.

\section{Renormalization in the presence of massive sl vector mesons}

\subsection{Remarks on scalar QED; induced interactions and counterterms}

As mentioned in the introduction SLFT differs both in its concepts as well as
in its calculational techniques from Lagrangian- (or Euclidean action)-based
quantization theories. In section 3 these differences played a role in the
solution of the Velo-Zwanziger causality conundrum and in the present section
they will be exemplified in a full QFT.

The simplest nontrivial model for illustrating these differences (for reasons
which will become clear in the sequel) is scalar QED.

Being an S-matrix-based QFT, the starting point of SLFT is an S-matrix which,
following Bogoliubov, is formally written as a time ordered product of an
interaction density $L(x,e)~$as in (\ref{B}). The construction of interacting
sl fields uses Bogoliubov's map (\ref{C}) which converts free fields into
their interacting counterparts whose large-time asymptotic behavior reproduces
the scattering amplitudes associated to $S.$ In this section we will be
exclusively interested in the S-matrix; localization properties of interacting
fields will be mentioned in section 6.

The simplest nontrivial model for which the preservation of positivity
requires the use of sl localized fields is \textit{scalar QED}. Here
"nontrivial" refers to the appearance of a $2^{nd}$ order induced term $A\cdot
A\varphi^{\ast}\varphi$ and a $4^{th}$ order counterterm $(\varphi^{\ast
}\varphi)^{2}$ which has no counterpart in spinorial QED.

The independence of the large-time LSZ limits of causally separable fields on
their original localization is the basis of the SLFT perturbation theory. The
following is in part a recollection of arguments presented in section 3.

In lowest order we may start with the pl interaction density $L^{P}=A_{\mu
}^{P}j^{\mu},~j^{\mu}=i\varphi^{\ast}\overleftrightarrow{\partial^{\mu}%
}\varphi$ and use the linear relation with its short distance improved sl
counterpart and its escort $A_{\mu}^{P}=A_{\mu}-\partial_{\mu}\phi~$to write
\begin{align}
L^{P}  &  =L-\partial^{\mu}V_{\mu},~with~V_{\mu}=j_{\mu}\phi\label{V}\\
S^{(1)}  &  =\int L^{P}(x)d^{4}x=\int L(x,e)d^{4}x
\end{align}
This solves two problems in one stroke, the highest short distance
contribution to $L^{P}$ has been encoded into a divergence which drops out in
the adiabatic limit (second line) so that S is string-independent (the left
hand side) as well as renormalizable (the right hand side).

Actually one may forget the $L^{P}$ and formulate the SLFT construction solely
in terms of a $L,V_{\mu}$ pair fulfilling the $L-\partial V=0$ pair condition.
It turns out that the $L,V_{\mu}$ pair corresponding to vector mesons
interacting with lower spin particles and possibly among themselves is
uniquely determined: the interaction is completely determined in terms of its
particle content! In other words the $L^{P}$ can be defined in terms a SLFT
pair which is in turn defined of the particle content.

For the formulation of the higher order pair property it is convenient to use
the differential form of the pair property (as in section 3) and write
$d_{e}(L-\partial^{\mu}V_{\mu})=0.$ Further simplification is obtained by
using the "$Q$-formalism" with%
\begin{align}
Q_{\mu}  &  =d_{e}V_{\mu}=j_{\mu}u,~u=d_{e}\phi\\
&  d_{e}L-\partial^{\mu}Q_{\mu}=0\nonumber
\end{align}
The $Q_{\mu}$ turns out to have a better $m\rightarrow0$ behavior; it is only
logarithmically divergent and $\partial^{\mu}Q_{\mu}$ remains finite. The
logarithmic infrared divergence is a perturbative spacetime indication that
the massless limit of vacuum expectation values cannot be described in terms
of Wigner-Fock space which is simply a tensor product of a helicity space of
photons with that of charge-carrying Wigner particles \cite{Bu}. In other
words the S-matrix based perturbative SLFT formalism indicates that its
massless limits needs a (presently unknown) extended formulation of scattering theory.

A notational simplification for the higher order pair conditions is obtained
by using lightlike strings (not possible for $m=0$). In this case the massive
sl fields are functions in $e$ rather than distributions and hence all $e$ may
be set equal. The second order pair condition reads%

\begin{equation}
d_{e}TLL^{\prime}-\partial^{\mu}TQ_{\mu}L^{\prime}-\partial^{\prime\mu
}TLQ_{\mu}^{\prime}=0 \label{LQ}%
\end{equation}
and the extension to higher than second order is straightforward. They have no
counterpart in the standard pl setting and account for the strength of SLFT as
compared to the standard pl perturbation theory.

Violations of these relations are referred to as \textit{obstructions}. The
Bogoliubov S-matrix formalism is preserved by encoding these obstructions into
a redefinition of the interaction density $L\rightarrow L_{tot}=L+L_{2}%
+..$just as it was done for external potential interactions in section 3.

The time-ordering $T$ which fulfills (\ref{LQ}) is not necessarily the
"kinematic" time ordering~$T_{0}$ which is defined by attaching a
$-i2\pi(p^{2}-m^{2}-i\varepsilon)$ denominator to the momentum space 2-ptfct.
The scaling rule of renormalization requires that $T$ and $T_{0}$ share the
same scaling degree which in the presence of two derivatives leaves a
normalization freedom%

\begin{equation}
\left\langle T\partial_{\mu}\varphi^{\ast}\partial_{\nu}^{\prime}%
\varphi^{\prime}\right\rangle =\left\langle T_{0}\partial_{\mu}\varphi^{\ast
}\partial_{\nu}^{\prime}\varphi^{\prime}\right\rangle +icg_{\mu\nu}%
\delta(x-x^{\prime}) \label{T}%
\end{equation}
which leads to%
\begin{equation}
\partial^{\mu}\left\langle T\partial_{\mu}\varphi^{\ast}\partial_{\nu}%
^{\prime}\varphi^{\prime}\right\rangle -\left\langle T\partial^{\mu}%
\partial_{\mu}\varphi^{\ast}\partial_{\nu}^{\prime}\varphi^{\prime
}\right\rangle =i(1+c)\partial_{\nu}^{\prime}\delta(x-x^{\prime}) \label{U}%
\end{equation}
with an initially undetermined $c.$

The fulfillment of the second order pair requirement (\ref{LQ}) in the tree
approximation fixes $c=-1.$ The action of $S$ on one-particles states as the
identity operator $S\left\vert p\right\rangle =\left\vert p\right\rangle
~$takes care of the contribution from 2 contractions. The change of $T_{0}~$to
$T$ in $Tj_{\mu}j_{\nu}^{\prime}=T_{0}j_{\mu}j_{\nu}^{\prime}-g_{\mu\nu}%
\delta$ in all $Tj_{\mu}j_{\nu}~$accounts for the occurrence of the second
order induced $A_{\mu}A^{\mu}\left\vert \varphi\right\vert ^{2}$ term%

\begin{align}
TLL^{\prime}  &  =T_{0}LL^{\prime}-i\delta(x-x^{\prime})L_{2}\label{L1}\\
L_{2}  &  =gA_{\mu}A^{\mu}\left\vert \varphi\right\vert ^{2} \label{L2}%
\end{align}
which is usually attributed to the implementation of gauge symmetry, but here
it follows from the causality and positivity principle of interpolating fields
which guaranties the $e$-independence of $S$.

The reason for using the kinematical time-ordering $T_{0}$ instead of $T$ is
the comparison with GT. In SLFT it is more natural to use $T$ in which case
the second order $L_{2}$ remains encoded in $TLL^{\prime}.$

As already pointed out in section 3, GT has a \textit{formally} similar
structure. This is most clearly visible in a setting of gauge theory which
avoids the standard Lagrangian quantization of gauge theory (for spins
$s\geq1$ see \cite{Fro}) in favor of a perturbative S-matrix formulation as in
\cite{Scharf} \cite{Aste}. The physical $A_{\mu}(x,e)$ and its escort
$\phi(x,e)$ correspond to the gauge potential $A_{\mu}^{K}$ and the
St\"{u}ckelberg field $\phi^{K}$ acting in a ghost extended Krein space; the
authors show that $A^{K}-\phi^{K}$ has properties expected from $A_{\mu}^{P}.$

The two S-matrix-based constructions share the same improved short distance
behavior, but they achieve this in a very different way. Whereas in gauge
theory this is the result of enforced compensations between positive and
negative probabilities in intermediate states, the ultraviolet improvement in
SLFT accomplishes this by lessening the tightness of causal localization (but
not abandoning it !) and in this way reducing the strength of vacuum
polarization which is the only physical way to describe particles in terms of
physical (i.e. not gauge) interpolating fields.

Though both SLFT and gauge theory have the same short distance dimensions and
probably even share their Callen-Symanzik equations (and the related
asymptotic freedom property encoded in the beta-function), gauge theory
\textit{cannot account for the physics at finite distances let alone at long
distances}; infrared properties and the problem of confinement remain outside
its physical range.

Last not least the functional-analytic and operator-algebraic methods used in
deriving nonperturbative theorems from basic principles are not available in
Krein spaces. For this reason gauge theory is shunned in books addressing the
conceptual structure of QFT \cite{St-Wi}\cite{Haag}. The perturbative gauge
theoretic construction of a unitary S-matrix reveals this tension between
conceptual clarity and the efficiency of calculations which account for
experimental observations; it is a blessing for the impressive achievements of
the Standard Model but a curse for a on the principles of positivity and
causal localizability formulated QFT of the books.

Considering these conceptual deficiencies the perturbative calculations of a
gauge-invariant S-matrix of the Standard Model is a truly impressive
achievement. The idea that it represents a successful placeholder of an
unknown QFT is quite old and there have been many failed attempts to find the
real thing. The close formal analogy between gauge theory and SLFT suggest
that both may even exist side by side in a Krein extended Wigner-Fock space
containing additional indefinite metric degrees of freedom \footnote{J. Mund
seems to have discovered such a "hybrid" formulation (private communication).}.

Presently there exist no higher than second order SLFT calculation. Higher
order loop calculations in SLFT are much more laborious than calculations in
gauge theory. The gauge theoretic $4^{th}$ order calculation establishes the
existence of a $c(\varphi^{\ast}\varphi)^{2}~$counterterm. In contrast to the
second order $A\cdot A\varphi^{\ast}\varphi$ contribution its strength $c$ is
a new parameter which is not determined by electromagnetism of the e.g.
$\pi^{+}~$meson\footnote{Using such a model to describe electromagnetic
interactions of charge-carrying pions one usually sets $c=0.$}.

Could this counterterm in GT be an induced contribution in SLFT ? This
question is not as crazy as it sounds. The above $L^{P}$ theory is by itself
nonrenormalizable; its short distance dimensions and the number of
counterterms increase with perturbative order. Yet if "guided" in the above
sense by a $L,V_{\mu}$ pair it shares the finite number of possible free
varying parameters with the SLFT $L$ description.

The still missing answers to such questions are not only owed to the fact that
the number of theoreticians who are presently working on SLFT problems can be
counted on one hand but they also find their explanation in that the necessary
calculations are more involved than those based on pl fields. The sl setting
of QFT is the only known way to uphold the principles of QFT for \textit{all} fields.

The SLFT approach also touches on an old mathematical problems which arose
from QFT in the late 60's. The question was whether fields with $d_{sd}%
=\infty$ (polynomial unbounded) fields as e.g. Wick-ordered exponential
functions of pl fields as $\exp g\varphi$ have a well-defined mathematical
status. This led Jaffe to extend the notion of Schwartz distributions to a
general class of distributions which still allows smearing with a dense set of
compact localized Schwartz test functions.

The SLFT guided construction of the $L^{P}~$pl setting requires to identify pl
observables in both settings and suggests to identify the interpolating state
creating fields of the $L^{P}$ theory with Jaffe fields. They correspond to
the well-behaved sl interpolating fields: the two theories share not only the
S-matrix but also their local observables whereas the states in the $L^{P}$
theory remain singular in the sense of Jaffe. Such singular fields are not
required to have the usual domain properties which one needs to generate
operator algebras from fields so that the algebraic localization of compact
spacetime regions is fully accounted for by observables.

After having exemplified the main difference between gauge theory and SLFT in
the model of scalar QED, the following subsections will present low order
calculations in other models in which vector mesons couple to lower spin
matter fields and among themselves. This includes in particular the Higgs
models for which, different from the standard treatment, the form of the
Mexican hat potential and its spontaneous symmetry breaking is not imposed but
rather \textit{induced} as a consequence of $e$-independence of $S.$ Even more
surprising is that the division into observables and sl interpolating is very
different from what one naively expects: neither the field strength $F_{\mu
\nu}$ nor the Higgs field is a pl observable.

\subsection{The perturbative S-matrix in the SLFT setting}

The appropriate formalism for the direct perturbative calculation of the
on-shell $S$-matrix is based on the adiabatic limit of Bogoliubov's
operator-valued time-ordered $S(g)$ functional. Its adjustment to SLFT has
been mentioned in (\ref{B}) in section 3 and further explored in the previous subsection.

Time-ordering of quantum fields mathematically represented by operator-valued
distributions is characterized in terms of properties among which the causal
factorization is the physically most important one. The Epstein-Glaser
formalism \cite{E-G} provides a perturbative computational scheme in which the
time-ordering of $n+1$ pl interaction densities is inductively determined in
terms of the $n^{th}$ ordered time-ordered product. The formulation in the
presence of sl fields is more involved and has not been carried out beyond
second order. Preliminary results reveal that a systematic $n^{th}$ order
construction requires the use of new concepts \cite{CMV}.

The E-G perturbation theory for the S-matrix can be extended to sl fields
(\ref{C}). The result is a formula which maps a field in the local equivalence
class of Wick-ordered composites of free fields into the equivalence class of
"normal ordered" relative local interacting fields which act in the same
Wigner-Fock Hilbert space but are nonlocal with respect to their free
counterparts. Nowhere does this formalism refer to Lagrangian quantization.
For gauge theory this was first carried out in \cite{Du-Fr} where it was shown
the time-ordering of the S-matrix passes to that of retarded products in terms
of fields.

The power-counting restriction of renormalizability $d_{sd}(L^{P})\leq4$ is
violated if one of the spin/helicity of the particles is $\geq1.$ For
interactions involving particles with highest spin $s=1$ the $d_{sd}(L^{P})=5$
there are two ways to recover renormalizability. Either by converting $L^{P}$
into a "gauge pair" $L^{K},V_{\mu}^{K}$ which requires the extension of the
Wigner-Fock space by indefinite metric degrees of freedom and possibly BRST
ghosts, or one maintains the physical degrees of freedom (and with it the
positivity of the Wigner-Fock Hilbert space) by converting $L^{P}$ into sl
$L,V_{\mu}$ pair.

For the rest of the paper we will stay with models which are sl renormalizable
$d_{sd}(L)\leq4.$ This includes all couplings whose particle content consists
of $s=1$ coupled to $s<1~$and among themselves. In the case of massless sl
vector potentials the escort $\phi$ diverges as $m^{-1}$ and the large-time
LSZ derivation of the S-matrix breaks down (the on-shell restrictions of
correlations develop logarithmic $m\rightarrow0$ singularities) and with it
the S-matrix based SLFT construction.

However some remnants of the SLFT construction can be saved; the exact
one-form $d_{e}\phi$ and hence also the $Q_{\mu}=d_{e}V_{\mu}~$is only
logarithmically divergent and $\partial^{\mu}Q_{\mu}$ remains convergent.
Hence even in case of breakdown of the S-matrix as a result of infrared
problems the $L,Q_{\mu}$ pair condition%
\begin{equation}
d_{e}L-\partial_{\mu}Q^{\mu}=0,\text{ }Q_{\mu}=d_{e}V_{\mu} \label{st}%
\end{equation}
remains a nontrivial condition. In fact it is this weaker formulation of
$e$-independence which corresponds to the BRST invariance of gauge theory.

In the previous subsection it was shown that, although the second order pair
condition in its original form is violated, it is possible to encode the
obstructing contribution $L_{2}$ into a redefinition of the interaction
density. It is helpful to formulate this idea in a model-independent way.

The definition of second order obstruction against the naive form of the
$L,Q_{\mu}~$pair property reads (using lightlike $e^{\prime}s$ which can be
identified)
\begin{align}
O^{(2)}  &  :=d_{e}TLL^{\prime}=T\partial^{\mu}Q_{\mu}L^{\prime}-\partial
^{\mu}TQ_{\mu}L^{\prime}+TL\partial^{\prime\mu}Q_{\mu}^{\prime}-\partial
^{\prime\mu}TLQ_{\mu}^{\prime}\label{d}\\
O^{(2)}  &  =\delta(x-x^{\prime})d_{e}L_{2}(x,e)\nonumber
\end{align}
Encoding them into interaction density one obtains%
\begin{equation}
L_{tot}:=L+gL_{2},~~S(g)=T\exp\int ig(x)L_{tot}(x,e)d^{4}x \label{ob2}%
\end{equation}

This change of bookkeeping which converts higher order obstruction into
induced contributions $L_{n}~$amounts$~$a change of $L\rightarrow L_{tot}$ in
the Bogoliubov $S(g)$~is important. It affects the higher orders; the third
order obstruction is now
\begin{equation}
O^{(3)}(g,g,g)=d_{e}\left[  TL(g)L_{2}(g^{2})+\frac{i}{3}TL(g)L(g)L(g)\right]
\label{ob3}%
\end{equation}
In models of interacting $s=1~$vector mesons as the Higgs model or scalar
massive QED the third order obstruction vanishes in the adiabatic limit and
the induced contributions account for the Mexican hat potential. As a
consequence the terms in this potential are induced and not postulated for the
purpose of implementing SSB. This will be explicitly verified in the following subsections.

The $L,Q_{\mu}~$pair condition and its higher order extension within the sl
Bogoliubov-Epstein-Glaser setting is also meaningful for $d_{sd}(L)>4.$ The
before mentioned "minimal" models contain only induced contributions but their
number increases with the perturbative order. By definition of minimal there
are no higher order counterterm parameters so the model depends only on those
parameters which are already present in the interaction density $L.$

The conceptual and mathematical superior aspects of SLFT poses the question
whether it is possible to pass directly from SLFT to pl ultra fields, thus
avoiding the pl counterterm formalism. This problem will come up again in
connection with cubic $h=2$ selfinteractions in the next section (\ref{rel}).

\subsection{External source models}

Consider a vector potential coupled to an conserved classical current $j_{\mu
}$\ \cite{MRS2}. The S-matrix and the interacting vector potential are%
\begin{align}
&  L^{P}=A_{\mu}^{P}j^{\mu}=A_{\mu}j^{\mu}-\partial_{\mu}(\phi j^{\mu
}),~hence~L=A_{\mu}j^{\mu},\ V_{\mu}=\phi j_{\mu}\label{s}\\
&  S_{e}(g)=T\exp i\int g(x)L(x,e)\overset{g(x)\rightarrow g}{\rightarrow
}S=\exp ig\int\int j^{\mu}i\Delta_{F}j_{\mu}^{\prime}:\exp ig\int A_{\mu}%
^{P}j^{\mu}:\nonumber\\
&  A_{\mu}^{ret}(x,e,e^{\prime})=S_{e}{}^{-1}(g)\frac{-i\delta}{\delta f_{\mu
}(x,e^{\prime})}S_{e}(g,j\rightarrow j+f)|_{f=0}=A_{\mu}(x,e)+\int G_{\mu
\mu^{\prime}}^{ret}j^{\mu^{\prime}}\nonumber\\
~  &  G_{\mu\mu^{\prime}}^{ret}(x,e;x^{\prime},e^{\prime})=(-\eta_{\mu
\mu^{\prime}}-\partial_{\mu}e_{\mu^{\prime}}I_{e^{\prime}}+\partial
_{\mu^{\prime}}e_{\mu}^{\prime}I_{-e}+(ee^{\prime})\partial_{\mu}\partial
_{\mu^{\prime}}I_{e}I_{-e^{\prime}})G^{ret}(x-x^{\prime})\nonumber
\end{align}
The direct use of $L^{P}$ with $d_{sd}(L^{P})=d_{sd}(A_{\mu}^{P})=2$ leads to
a delta function ambiguity $g_{\mu\nu}c\delta(x-x^{\prime})~$in the
time-ordered $A_{\mu}^{P}$ propagator which accounts for a replacement
$i\Delta_{F}\rightarrow i\Delta_{F}+\frac{c}{m^{2}}\delta$ in the second line.
This in the pl formulation undetermined counterterm renormalization parameter
in the S-matrix and in $A_{\mu}^{P,ret}$ is absent in the less singular sl formulation.

In that case $S$ is independent of $c$ and (by use of current conservation)
the interacting field does not depended on $e^{\prime}.$ As expected the field
strength remains pl. Hence the avoidance of the direct use of $A_{\mu}^{P}$ in
the calculation maintains the predictive power of the model. If needed one can
convert the sl setting with the help of $\phi(x,e)$ to a $A_{\mu}^{P}$. In
contrast to the directly calculated $A_{\mu}^{P}$ this via sl determined pl
potential is "better".

Passing from external source to \textit{external potential} problems the
differences between the direct pl results and those obtained via the sl detour
are much stronger (section 3).

\subsection{Hermitian $H~$coupled to a massive vector potential}

The coupling of a vector potential to a Hermitian scalar matter field $H$
comes with a new phenomenon. In addition to a change of the time-ordered
product of the $H$-field there is now a genuine \textit{induction}~of $H$-selfinteractions.

The "germ" of an interaction density (the "ignition") for an $A_{\mu},H$ field
content is the $mA\cdot AH$ coupling, where the vector meson mass factor $m$
accounts for the classical dimension $d_{eng}=4$ and also indicates that the
model has no nontrivial Maxwell limit (the reason why it was discovered a long
time after QED)$.$ Its sl operator dimension is $d_{sd}=3,$ hence the germ is
a superrenormalizable sl interaction density. The first order $L,Q_{\mu}$ pair
property ($Q_{\mu}=d_{e}V_{\mu}$) requires the presence of the escort $\phi$
also in $L~$and leads to ($L,Q_{\mu}$ relation easy to check) $\ $
\begin{align}
&  L=m\left\{  A\cdot(AH+\phi\overleftrightarrow{\partial}H)-\frac{m_{H}^{2}%
}{2}\phi^{2}H\right\}  +U(H),~U(H)=mc_{1}H^{3}+c_{2}H^{4}\nonumber\\
&  V_{\mu}=A_{\mu}\phi H+\frac{1}{2}\phi^{2}\overleftrightarrow{\partial_{\mu
}}H,~~\ d_{e}(L-\partial^{\mu}V_{\mu})=0,\text{ }L^{P}=L-\partial^{\mu}V_{\mu
}. \label{first}%
\end{align}
A systematic determination of this first order pair $L,V_{\mu}$ pair starting
from the simplest coupling (the "germ") $gmA\cdot AH~$of the $A$-$H$ particle
content and a general ansatz for $L$ and $V_{\mu}$ containing all
kinematically possible $d_{sd}\leq4~$terms$~$(19 terms in $L$) which can be
formed from $H,A_{\mu}$ and its escort $\phi$ shows that (\ref{first}) is (up
to $\partial_{\mu}~$divergence terms and exact $d_{e}$ differentials) is the
unique solution \cite{MS}. However a verification that the $L,V_{\mu}$ pair
satisfies the pair condition requires only the use of free field equations and
the relations between $A_{\mu}$ and its escort and will be left to the reader.

The first order pair condition does not determine the strength of the
$H$-selfinteractions since $e$-independent contributions to $L$ simply pass
through the pair condition. The necessity of their presence which includes the
determination of the $c_{i}~$in (\ref{first}) is seen in second and third
order. This "induction" of additional contributions with well-defined
numerical coefficients is a new phenomenon of SLFT; there is a formal
similarity with the imposition of the second order BRST gauge invariance on
the $S$-matrix \cite{Scharf} but the essential difference is that the
$e$-independence of $S$ is a consequence of the positivity and causal
localization principle of QFT.

For the S-matrix one only needs the second order tree component to the
obstruction $O^{(2)}$ in$~$(\ref{d}) In addition to a second order change of
the time ordering of the propagator involving derivatives of $H~$which
parallels that in (\ref{T})~one now encounters a genuine second order
induction (\ref{ob2})
\begin{equation}
L_{2}=g[(m_{H}^{2}+3c_{1}m^{2})H^{2}\phi^{2}-\frac{m_{H}^{2}}{4}\phi^{4}%
+c_{2}H^{4}] \label{two}%
\end{equation}
Finally the vanishing of the third order tree contribution fixes the values of
$c_{1},c_{2}$ in terms of the three physical parameters of its field content
which were already present in the germ namely $g,m,m_{H}.$ To allow for a
comparison with the Higgs mechanism we write the result in the form
\begin{equation}
L_{tot}^{(2)}=mA\cdot(AH+\phi\overleftrightarrow{\partial}H)-V(H,\phi
),~V=g\frac{m_{H}^{2}}{8m^{2}}(H^{2}+m^{2}\phi^{2}+\frac{2m}{g}H)^{2}%
-\frac{m_{H}^{2}}{2g}H^{2}\quad\label{pot}%
\end{equation}
where $L_{tot}^{(2)}=L+\frac{g}{2}L_{2}.$ The appearance of a quadratic mass
term is the result of writing the interaction density as if it would be part
of a classical Lagrangian of gauge potentials. The reader may fill in the
details of the straightforward calculations by himself or look up the more
detailed presentation in \cite{MS}.

Apart from a mass contribution the $V~$looks like a field-shifted Mexican hat
potential. \textit{But different from the Higgs mechanism it has not been
obtained by postulating a Mexican hat potential and subjecting it to a shift
in field space}. It is rather\textit{ induced} by a renormalizable
$A,H$\textit{ field content }and it is the \textit{unique renormalizable QFT
with this field content}. There is simply no room for imposing a Mexican hat
potentials since the induction of the $H$ and $\phi$ selfinteractions is a
consequence of $e$-independence of the S-matrix which in turn is a consequence
of scattering theory involving $d_{sd}=1$ causally separable space- or
light-like strings.

The SSB picture of the Higgs model also reveals another common
misunderstanding, this time about SSB. The Mexican hat potential together with
the shift in field space is \textit{not the definition} of SSB but\textit{
rather a way to implement} such a situation \textit{when it is possible}. The
definition of SSB is rather the \textit{existence of a locally conserved
current whose global charge diverges}. This is only possible in the presence
of massless Goldstone bosons and all verbal attempts to make SSB consistent
with a mass gap (a photon becoming fattened to a vector meson by eating a
Goldstone) only obscure the interesting correct understanding.

\textit{QFT is not a theory which can create the masses of its model-defining
field content}. In particular SSB is not about creating finite masses from an
initially massless situation; to the contrary it is about how to place a
massless particle (the Goldstone boson)\ into an interaction density so that
the current conservation remains that of a symmetric theory but some local
charges are prevented to converge in the infinite volume limit to a finite
global charge (the definition of SSB). The only known The prescription of a
field shift on a Mexican hat potential as the "Higgs mechanism" has to be seen
in a historical context; it helped to overcome the formal problems which one
faces when one tries to extend Lagrangian quantization from Maxwell's theory
of charge-carrying fields to a situation in which a vector potential couples
to a Hermitian matter fields. There are numerous historical illustrations of
situations for which important discoveries were made through formal
manipulations which were later replaced by a derivation which is consistent
with the principles of QFT. Incorrect placeholder are useful but only up to
the discovery of the real reasons.

A model of QFT is uniquely fixed in terms of its field content. The SLFT
setting (which seems to be the only one consistent with all principles of QFT)
for a $A_{\mu},H$ field content starts with a $A_{\mu}A^{\mu}H$ as the
simplest coupling and the rest is done by induction using the $L.Q_{\mu}$ pair
property which converts the heuristic physical content of the ill-defined pl
interaction density into the physically superior SLFT setting where the
"induction" resulting from the implementation of the pair property to all
orders unfolds the full content of SLFT.

\subsection{Selfinteracting vector mesons}

It is straightforward to check that there is no renormalizable $L,Q_{\mu}$
pair for a self-coupled singlet $(A\cdot A)^{2}.$ The principles of QFT as
embodied into the pair condition admit however selfinteractions between
multiplets ("colored") of vector potentials while imposing strong restrictions
on the "multi-colored" coupling parameters. In this case the germ is a $FAA$
selfinteraction and the general ansatz for the construction of a $L,V_{\mu}%
~$pair which includes the "colored" escorts is of the form
\begin{equation}
L=\sum(f_{abc}F_{c}^{\mu\nu}A_{a,\mu}A_{b,\nu}+h_{abcd}A_{a,\mu}A_{b,\nu}%
A_{c}^{\mu}A_{d}^{\nu})+\,\hbox{terms in}\,A_{a,\mu}\,\hbox{and}\,\phi
_{a}^{\prime}s
\end{equation}
where the couplings and the masses of the vector mesons are initially freely
variable parameters but, as expected, the first and second order pair
condition places strong restrictions on them \cite{MS}, among other things the
$f$ and $h$ are interrelated in the same way (Jacobi identities of reductive
Lie-algebras) as in gauge theory \cite{Scharf}; in particular the $A$-$\phi$
and $\phi$-$\phi$ couplings depend also on the masses of the vector mesons.
The main distinction to gauge theory is that these properties are direct
consequences of the principles of QFT and do not arise in the course of the
gauge theoretic extraction of physics from a unphysical (positivity-violating)
description through the imposition of gauge invariance.

The most interesting aspect of the SLFT formulation is that there remains a
renormalizability destroying second order induced selfinteraction which, if
left uncompensated, destroys renormalizability even though the interaction
density fulfills the power counting restriction of renormalizability. The way
to overcome this is to compensate this $d_{sd}=5~$term with a second order
contribution from a $A$-$H$ interaction with a scalar Higgs field\footnote{A
$s\geq1$ field would worsen the second order short distance behavior.}. This
is a totally different situation from the abelian $A$-$H$ interaction for
which such all second order terms stay within the power counting bound.
Neither case bares any physical resemblance to spontaneous symmetry breaking
since in both cases the field shifted Mexican hat potential is second order induced.

The idea of short distance compensations between contributions from different
spins arose in connection with supersymmetry. Although not invented for this
purpose, SUSY does improve the short distance behavior somewhat but not enough
to guaranty the renormalizability and preservation of supersymmetry in higher
orders. The situation of selfinteracting vector mesons is different, in that
the preservation of renormalizability is the raison d'\^{e}tre for the $H.$
Nature does not have to decide between a symmetry and its SSB, rather the
existence of the $H$ is directly connected to the preservation of its
positivity and causality principles or in other words a massive $A_{\alpha
}^{\mu}$ field content by itself is not consistent.

Gauge symmetry is not a physical symmetry so there is nothing to break; all
these physically incorrect pictures evaporate if one maintains causality
\textit{and} positivity which is perturbation theory is only possible by
starting with an sl $L,V_{\mu}$ or $Q_{\mu}~$pair property. The fibre bundle
like Lie structure of the $f_{abc}$ couplings is not the result of an imposed
symmetry it rather arises from the string-independence of the S-matrix which
in turn is a result of LSZ scattering theory of interacting causally separable
positivity obeying quantum fields; hence the situation is very different from
the superselection structure of unitary representation classes of observable
algebras which leads to the notion of inner symmetries. This shows that
quantum causality is much more fundamental than its classical
Faraday-Maxwell-Einstein counterpart.

Having thus strengthened the conceptual understanding of interactions between
vector mesons in the Standard Model one may ask whether SLFT contains also
messages about their coupling to matter. In recent work \cite{GMV} it was
shown that SLFT does not only restrict the selfcouplings between vector mesons
and requires the presence of a Higgs particle in the presence of
selfinteracting massive mesons but it also restricts their coupling to the
Fermion currents and their chirality properties. This is of particular
interests for massive $W_{\pm}.Z$ vector mesons and the photon, a\ case for
which the authors explain the restrictions from SLFT in detail.

\subsection{The pair condition for higher spins}

The extension of SLFT S-matrix construction to that of interacting higher
spins $s\geq2~$is an important issue about which one presently knows little.
There have been quite extensive investigations in a gauge theoretic equivalent
of the pair condition by Scharf \cite{Scharf}. In view of formal similarities
with SLFT it is interesting to take a closer look at some of his results.

Scharf looked at the simplest $s=2~$selfinteraction which is of a cubic form
$trh^{3}$ where $h_{\mu\nu}$ is the $s=2$ massless tensor field. The physical
interest in this model is connected with the use of $h_{\mu\nu}~$as a linear
approximation of the gravitational $g_{\mu\nu}~$field. As in SLFT, the short
distance dimension of integer spin gauge fields is equal to their classical
dimension in terms of mass units namely $d_{sd}=1.$In \cite{Scharf}~it was
shown that there exists no gauge theoretic trilinear selfinteraction $L^{K}$
with $d_{sd}(L^{K})=3$ without involving derivatives of $h_{\mu\nu},$ its
trace $h_{\mu}^{\mu}~$as well as ghost fields and their anti-ghost. He found a
cubic interaction density of $d_{sd}(L^{K})=5$ which is above the
power-counting bound of renormalization, but still presents a huge reduction
from the $d_{sd}(L^{P})=11$.

Taking into account that gravitational coupling carries a dimension and
expanding the Einstein-Hilbert Lagrangian in a suitable way using
$\kappa=\sqrt{32\pi G}$ as an expansion parameter, he arrived at a formal
connection of the classical expansion with the quantum-induced correction up
to second order; this was later extended to all tree orders \cite{Scharf,Du}.
The agreement of tree approximations with classical perturbation theory is not
unexpected in itself, but in the present context it relates two competing
ideas, one being of classical geometric origin (the Einstein-Hilbert action)
and the other the gauge theory of selfinteracting $h=2$ particles .

Christian Ga\ss \ showed recently (private communication) that SLFT provides a
simpler version of such a cubic selfinteractions in the form%
\begin{align}
L  &  =\kappa(2\partial_{\rho}h^{\kappa\lambda}\partial_{\sigma}%
h_{\kappa\lambda}+4\partial_{\beta}h_{\rho}^{\alpha}\partial_{\alpha}%
h_{\sigma}^{\beta})h^{\rho\sigma},~~h_{\mu\nu}:=A_{\mu\nu}^{(2)}\label{L}\\
&  d_{e}A_{\mu\nu}^{(2)}=\partial_{\mu}a_{\nu}+\partial_{\nu}a_{\mu},~
\end{align}
where $h_{\mu\nu}=A_{\mu\nu}^{(2)}$ is the sl helicity$~$2 potential from
(\ref{E}) section 2 (which already played a role in solving the D-V-Z
discontinuity problem \cite{MRS2}). Using the relation between $d_{e}$ and
$\partial_{\mu}$ of the second line one easily verifies that $d_{e}L$ is of
the form $\partial^{\mu}Q_{\mu}$ i.e. the above $L$ belongs to a $L,Q_{\mu}%
~$pair. Since massless $h\geq1$ fields are intrinsically sl, the corresponding
minimal models are expected to be "ultra-distributions" which are localizable
in spacelike cones.

For $h=1$ there exists no colorless selfinteraction whereas for $h=2$ the
situation seems to be reverse since the existence of colored cubic
selfinteractions can be excluded \cite{Bou}. A proof based on Scharf's
S-matrix gauge formalism can be found in \cite{Grig}.

The fact that there are no \textit{renormalizable} $s=2$ selfcouplings does
not exclude the possibility to find sl $L,Q_{\mu}$ pairs of interactions
between sl $h_{\mu\nu}$ with lower spin fields as $H$ or/and $A_{\mu}$. An
ansatz for $L$ which generalizes the $A_{\mu},H$ particle content of the
abelian Higgs model would be of the form ($h_{\mu\nu}$ massive)
\begin{equation}
L=mgh_{\mu\nu}h^{\mu\nu}H+\,U(H,h,\phi) \label{in}%
\end{equation}
where the first term represents the "ignition" i.e. the simplest
renormalizable ($d_{sd}=3$) interaction associated with a $h,H$ particle
content and $U$ contains all the remaining possible at most quadrilinear
couplings between $h_{\mu\nu},$ its 5 escorts $\phi_{\mu},\phi$ and $H$. Their
coupling strengths are determined from the first or second order ("induction")
pair condition.

The $L,Q_{\mu}$ pair may be uniquely determined, but it is rather improbable
that $d_{sd}(L)\leq4.$ It would be premature to dismiss $L,V_{\mu}$ pairs with
$d_{sd}(L)>4.$ The example of pl models with $d_{sd}(L^{P})=5,~$which in the
standard pl renormalization theory leads to a with perturbative order
increasing number of renormalization parameters but under the guidance of an
S-matrix-equivalent sl pair turns into an improved formalism. This upgraded
$L^{P}$ description contains $d_{sd}\rightarrow\infty$ pl fields but shares
its parameters, the S-matrix and its pl local observables with the SLFT
renormalization theory.

Presently our understanding of the consequences of the higher order SLFT pair
requirements is too scarce to say anything credible about $~L,V_{\mu}$ pairs
with $d_{sd}(L)>4$. A clarification of this important issue will be left to
future research.

\section{Dynamical string-localization of interacting fields}

Free massive pl fields can not only be converted into their sl counterparts by
integration along strings but the directional $e^{\mu}\partial_{\mu}$
differentiation on sl free fields permits also the return to its pl form.
Together with their Wick-ordered composites they form the local equivalence
(sl extended Borchers-) class $\mathcal{B}$ of free fields (pl fields are
viewed as special cases of sl).

Recall that for the construction of the S-matrix corresponding to a prescribed
particle content one uses pl fields for $s<1$ and those special $s\geq1$
massive sl potentials which were constructed in section 2.3 by "fattening"
their uniquely defined sl massless counterpart. Together with the uniquely
defined pl Proca potential and a scalar sl field referred to as the escort
they constitute a triple of relatively causally localized fields which act in
the massive $s=1$ Wigner-Fock space. They fulfill a linear relation which is
the basis for the construction of renormalizable sl interaction densities with
string-independent S-matrices.

This "kinematic" sl localization of $s\geq1~$free fields is important for the
construction of the S-matrix ala Bogoliubov. But it does not account for the
\textit{physical localization of the interacting fields} which is not in the
hands of the calculating physicists but is determined by the particle content
of the model. To distinguish between the two the localization of the
interacting fields will be referred to as "dynamic localization".

To understand this important point it is helpful to recall the form of the
\textit{Bogoliubov map} which relates the pl or sl Wick-ordered free fields
from the local equivalence class of free fields $\mathcal{B}$ to that of
normal ordered interacting fields $\mathcal{B}|_{L}$ (\ref{C}). For \thinspace
pl gauge theoretic interactions densities $L^{K}$ this problem has been
studied in \cite{Du-Fr}.

One important result is that this perturbatively defined linear Bogoliubov map
preserves the relative causality of fields but not the algebraic structure.
This is in agreement with algebraic QFT which is based on the idea that the
full physical content of QFT in the presence of interactions is contained in
the net of spacetime localized algebras \cite{Haag}. What is shared between
$\mathcal{B}$ and $\mathcal{B}|_{L}~$in case of massive vector potentials is
the Wigner-Fock Hilbert space in which these fields act.

This transfer of pl causality undergoes significant changes in the presence of
sl fields. As in the calculations in the previous section one uses a lightlike
$e$, in this case no directional testfunction smearing is necessary.

For the understanding of changes in localization caused by the Bogoliubov map
it is not necessary to enter the details of perturbative renormalization. It
suffices to understand the relations between free fields in $\mathcal{B}$
which result from the assumption that their interacting images of the
Bogoliubov map into the target spaces $\mathcal{B}|_{L_{tot}\text{ }}$and
$\mathcal{B}|_{L_{tot}^{P}\text{ }}$coalesce. Hence one may omit the
prefactors $S^{-1}$ in the Bogoliubov maps and write%

\begin{align}
&  S(g(x)L_{tot}^{P}+\lambda f\varphi_{g})|_{\lambda=0}\overset{a.l.}{\simeq
}S(g(x)L_{tot}+\lambda f\varphi)|_{\lambda=0}\label{rel}\\
&  \varphi_{g}|_{L_{tot}^{P}}=\varphi|_{L_{tot}},~~\varphi_{g}(x,e)=\varphi
(x,e)+\sum_{k=1}^{N}\varphi_{k}(x,e) \label{con}%
\end{align}
where the $\varphi_{g}|_{L_{tot}^{P}}$ refers to the interacting image of
$\varphi_{g}$ under the $L_{tot}^{P}$ Bogoliubov map.

The information about the localization of an interacting field $\varphi
|_{L_{tot}}$ is contained in the left hand side\footnote{The important point
is that the $L^{P}$ Bogoliubov map preserves localization; hence one can use
it to find out whether the image of $\varphi$ under $L~$is sl or pl.} whereas
its renormalizability status (finite or infinite $d_{sd}$) can be red off on
the the right hand side. Fields which are renormalizable and at the same time
pl in the $L^{P}$ setting represent observables whereas renormalizable fields
which are sl on the $L_{tot}^{P}$ side are sl interpolating fields.

The formal combined map of $\mathcal{B}$ into itself is highly non-linear and
generally changes localization properties; this is the price for the
preservation of renormalizability. The $\varphi_{k}(x,e)$ in (\ref{con}) are
determined by the induction
\begin{equation}
\varphi_{k+1}(x,e)=ig\int T(L_{tot}^{P}(x^{\prime})-L_{tot}(x^{\prime
},e))\varphi(x,e)=ig\int[\partial^{\prime\mu}T]V_{tot,\mu}(x^{\prime
},e)\varphi_{k}(x,g)d^{4}x^{\prime} \label{it}%
\end{equation}
where $[\partial^{\prime\mu},T]$ denotes the difference between the
$\partial~$acting outside and inside the time-ordering which either vanishes
or contributes a $\delta$-term.

In massive QED this conversion (\ref{con}) has no effect on pl observables;
fields as $A_{\mu}^{P}$ and$\ F=\operatorname{curl}A^{P}$ simply pass through
since with $V_{\mu}=\phi j_{\mu}$ and $\varphi_{0}=A_{\mu}^{P}~$the right hand
side (\ref{it}) vanishes and hence%

\begin{equation}
A_{\mu}^{P}(x)|_{L^{P}}=A_{\mu}^{P}(x)|_{L},~~F_{\mu\nu}|_{L^{P}}=F_{\mu\nu
}|_{L}\quad\label{ob}%
\end{equation}
The idea underlying such conversions was first used by Mund \cite{Mund} in the
context of massive spinor QED. He calculated higher orders for the
charge-carrying $\psi$ (spinor or complex scalar) and found consistency with
\begin{align}
&  \psi(x)|_{L^{P}}=e^{ig\phi(x,e)}\psi(x)|_{L}\label{Mund}\\
&  \psi(x)|_{L}=e^{-ig\phi(x,e)}\psi(x)|_{L^{P}}%
\end{align}
The formula is reminiscent of gauge transformation, however its physical
content is quite different.

A particularly interesting application of the conversion formalism arises in
the Higgs model. Different from massive QED, neither the $s=1$ field$\ A_{\mu
}^{P}|_{L^{P}}$, nor $H_{L^{P}}$ are local observables. Using the form of
$V_{\mu}~$in (\ref{first}) one finds that $H$ is transformed into $H_{1}$
\begin{equation}
H_{1}(x,e)=-\int[\partial^{\prime\mu},T]V_{\mu}(x^{\prime})H(x)d^{4}x^{\prime
}=\frac{1}{2}:\phi^{2}(x,e): \label{H}%
\end{equation}
i.e. $H$ is against naive expectations not an observable but rather represents
a sl interpolating field. The same holds for $A^{P}$~or its
$F=\operatorname{curl}A^{P}.$

Allowing additive composite modifications $H\rightarrow H+polyn(H,A^{P})$
which preserve the asymptotic scattering state of the $H$-particle does not
change the situation. The same holds for the $A_{\mu}^{P}$ or
$F=\operatorname{curl}A^{P}$ $\ $Hence both fields which are linearly related
to the particle content of the model are interpolating fields and do not
represent observables. The fact that $A_{\mu}^{P}|_{L^{P}}$ and $F$ are
observables\footnote{The line integral over an observable commutes with
"switching on" the interaction and does not represent an interpolating
fields.} in massive QED but not in the Higgs model shows that the
observable-interpolating field dichotomy is not a kinematic property.

Hence fields representing local observables in the Higgs model are necessarily
composite. A composite field which exists in every model is the interaction
density $L_{tot}^{P}|_{L_{tot}^{P}}=L_{tot}|_{L_{tot}}.$ The right hand side
was calculated in (\ref{pot}) and the computation of $L^{P}~$will be contained
in a forthcoming publication \cite{MS}.

A better understanding about the singular structure of pl fields may shed new
light on the localizability of $d_{sd}=\infty$ fields which arose in
connection with summing up graphical structures in certain nonrenormalizable
models \cite{Ba-S}. This problem was taken up by Arthur Jaffe \cite{Jaffe}%
\cite{Sol} who discovered a new class of singular distributions which still
permit smearing with a dense set of compact supported Schwartz testfunctions.
These Jaffe distributions had no impact on QFT because the unguided pl
nonrenormalizability with its infinite number of renormalization counterterm
parameters does not present a well-defined arena for physical applications.
Such $d_{sd}=\infty~$pl fields are probably too singular to generate operator
algebras, but they may still create physical states in the SLFT-guided $L^{P}$ formalism.

In section 3.2 we sketched the application of this formalism to interactions
with external potentials. Such interactions do not lead to loop contributions.
This simplicity of only induced contribution promises an interesting
mathematically controllable playground for the study of the pl localization
properties of observables and that of sl interpolating fields.

SLFT is presently the last step in a process of dissociating QFT from its
historic ties with Lagrangian quantization. When shortly after the discovery
of renormalized QED Arthur Wightman presented his "axiomatic" formulation of
QFT in terms of pl fields, it appeared to be the most appropriate intrinsic
formulation which can be extracted from Lagrangian quantization and Wigner's
representation theory \cite{St-Wi}. In his algebraic formulation of Local
Quantum Physics (LQP) Rudolf Haag proposed a setting of QFT based on a net of
localized operator\ algebras representing observables which removed the last
vestiges of quantization \cite{Haag}.

The next step was taken in the 80s by Buchholz and Fredenhagen who showed that
the existence of observable algebras and suitably defined particle states
guaranties the presence of operators localized in arbitrary narrow spacelike
wedges (whose cores are strings) which create these particle states from the
vacuum \cite{Bu-Fr}. These constructions were too far removed from the
exigencies of renormalized perturbation theory in order to have a direct
impact on calculations.

As a result it tooks more than 30 years to incorporate these observations into
a new sl perturbation theory in whose discovery the understanding of the
noncompact localization of Wigner's infinite spin matter was an important
catalyzer \cite{MSY2}. Fortunately one does not have to go through the details
of this history in order to do perturbative calculations. But what may be
interesting to note is that, different from Wightman's extraction of his
axiomatic setting from what one learned from the mathematically rather
ill-defined rules of Lagrangian quantization, the construction of SLFT took
the opposite path by converting ideas from LQP into perturbatively accessible computations.

Its most remarkable physical property is that observables are distinguished
from interpolating fields in terms of localization, which is of course to be
expected in a theory based on causal localization, but which GT could not accomplish.

There remains the question how GT with its lack of quantum positivity for
interpolating fields achieves to be such an amazingly successful description.
This will be commented on in the concluding remarks.
\begin{verbatim}
Erratum: the claim that the interacting Proca field $A_{\mu}^{P}$ of the abelian Higgs model is an sl interpolating
field is incorrect. The correct propagator is transverse which implies that $(A_{\mu}^{P})_{1}=0$ i.e. $A_{\mu}^{P}$ remains a pl
observable.  
\end{verbatim}

\section{Resum\'e, loose ends and an outlook}

SLFT is a formulation of QFT in which renormalizable interacting fields
maintain the tightest possible localization which is compatible with quantum
positivity and causality. In contrast to gauge theory its physical range is
not limited to local observables and the S-matrix but also includes
string-local interpolating fields which mediate between the causality
principles of QFT and the string-independent scattering properties of
particles. All degrees of freedom are provided by Wigner's particle
representation theory.

As described in the introduction the discovery of SLFT was triggered by the
construction of sl free fields associated to Wigner's positive energy infinite
spin representation \cite{MSY2}. Yngvason's 1970 No-Go theorem \cite{Y}
precluded the existence of pl fields. It turned out that Wigner's massless
infinite spin representation presents a much stronger barrier against pl
localization than that observed by Weinberg and Witten in massless finite
helicity representations. The Weinberg-Witten No-Go theorem excludes the
existence of conserved higher helicity pl currents and energy-momentum
tensors; in view of the absence of massless pl vector potentials and the fact
that the existence of pl massless limits depends on the short distance
dimension $d_{sd}$ this is hardly surprising $.$

The infinite spin case excludes the existence of pl composites; more general:
the causal localization of infinite spin matter is necessarily noncompact
\cite{LMR} in concordance with smearing sl fields with directionally compact
localized test functions $f(x,e),~e^{2}=-1.$ Closely related is that
\textit{infinite spin matter cannot interact with ordinary (finite spin)
quantum matter ,} but through its energy-momentum tensor its backreaction on
classical gravity may lead to a noncompact form of gravity. Quantum inertness
combined with gravitational reactivity are properties attributed to dark
matter \textit{\cite{dark}}. Since the sl infinite spin energy-momentum tensor
is known as a bilinear form \cite{PL} such a calculation appears feasible.

The existence of sl infinite spin field with \textit{finite} $d_{sd}$
suggested that the renormalizability destroying $d_{sd}=s+1$ increase of short
distance dimension can be avoided by using sl fields. This was the start for
the construction of sl potentials for finite $s,h$ which provided the
positivity preserving (the Gupta-Bleuler degrees of freedom avoiding)
$d_{sd}=1$ potentials. As mentioned in section 3 the absence of pl currents
does not exclude the existence of local charges which are localized in
arbitrary small spacetime regions.

The weakening of causal localization in SLFT should not be misunderstood as
(what is commonly referred to as) "nonlocal"\footnote{The authors of
\cite{MRS1} had problems with referees who rejected the work with the argument
that SLFT is nonlocal.}. The use of covariant semi-infinite space- or
light-like half-lines does not get into conflict with the causality
prerequisites of scattering theory (namely the possibility of placing an
arbitrary number of fields in relative spacelike positions), nor is the
derivation of important structural theorems (TCP, Spin\&Statistics) impeded.

Among the continuously many sl potentials only one for each $s$ plays a role
in SLFT perturbation theory. The key observation for its construction is that
the equation $\operatorname{curl}A=F$ for a \textit{sl massless field}
$A_{\mu}(x,e)$ \textit{acting in the Wigner-Fock helicity space} associated to
the ($m=0,h=\pm1$) Wigner representation has a unique solution which replaces
the positivity violating pl potential of GT.

By a process referred to as "fattening" (section 2) this solution selects
among the many possible massive sl potentials (which act in the $s=1$
Wigner-Fock Hilbert space of the unique pl Proca potential) a distinguished sl
vector potential. Together with a canonically constructed scalar sl potential
$\phi(x,e)$ (the escort) one obtains a triple of linear related fields
$A_{\mu}-\partial_{\mu}\phi=A_{\mu}^{P}$ which act in the $s=1$ Wigner Fock
space and belong to the linear part of the causal equivalence class of
(Wick-ordered) free fields associated to the Wigner representation $(m>0,s).$

The string independence expressed as the pair relation $d_{e}(A-\partial
\phi)=0$ is the basis for constructing a renormalizable interaction density
$L(x,e)$ which couples the $s=1~$sl $A$ and $\phi$ fields to lower spin free
fields which remain pl. Together with a suitably defined vector density
$V_{\mu}~$ one arrives at the pair relation $d_{e}(L-\partial V)=0$ which
insures the string-independence of the S-matrix which is obtained by taking
the adiabatic limit of time-ordered product of the interaction density. The
lowest order pair relation may need an extension by induced terms which result
from the implementation of higher order pair conditions. This is a new
phenomenon which has no counterpart in the old pl perturbation theory. \ 

The interacting quantum fields associated to this S-matrix are constructed in
terms of the Bogoliubov map which converts pl or sl fields from the causal
equivalence class of Wick-ordered free fields into their normal ordered
interacting counterpart. The restriction to pl $s<1$ and sl $s=1$ free fields
is only necessary in the construction of the S-matrix; the Bogoliubov map can
be applied to \textit{any} (pl or sl, elementary or composite) field in the
free field class.

Its interacting target fields have in general a different localization from
their source fields. The target localization has to be determined in the
$L^{P}$ setting (see previous section). Renormalizable ($d_{sd}<\infty$)
fields in the $L$ target space (independent of their pl or sl localization)
represent observables if their $L^{P}$ source fields are pl; otherwise they
represent interpolating sl fields.

SLFT has been applied up to second order to all models in which vector mesons
interact with themselves or with $s<1$ particles. The by far conceptually most
demanding and interesting QFT is the Higgs model which in its most simple
(abelian) form is the QFT in which a vector meson interacts with a $s=0$
Hermitian field. The first order pair turns out to be uniquely fixed and its
second order implementation induces a $H$ selfinteraction which looks as if it
would come from spontaneous symmetry breaking on a postulated Mexican hat
selfinteraction. The conceptual difference to SLFT is enormous.

A similar but somewhat more elaborate second order calculation for
selfinteracting massive vector mesons reveals that the coupling structure of
the \textit{leading} $d_{sd}=4$ \textit{contributions} up to second order is
that of a reductive Lie-algebra. The surprise is that, different from gauge
theory, this apparent symmetry in the $d_{sd}$ leading contribution has not
been imposed. In fact it is not even a symmetry in the sense in which this
terminology is used to describe unitary implemented inner symmetries.

Whereas symmetries and their spontaneous or complete breaking of
selfinteracting scalar particles can be freely imposed, there are strong
restrictions from first principles on the form of $s\geq1~$SLFT
selfinteractions which leave no such freedom; the form of selfinteractions in
the presence of $s\geq1~$is fully determined by the particle content of the
model and not at the disposition of the calculating theorist. The use of the
positivity violating gauge symmetry obscures this important insight. The
chirality theorem \cite{GMV} shows that these principles also affects the
coupling of selfinteracting vector potentials to Dirac fermions.

Another somewhat unexpected property is that renormalizable interaction sl
densities $L$ may produce second order sl $d_{sd}=5$ contributions which, if
left uncompensated, destroy the $e$-independence of $S$ as well as
renormalizabilty. The only way to save such a model is to enlarge its particle
content by a $A$-$H~$interaction which induces a compensating second order $A$
selfinteraction. \textit{This, and not SSB, is the raison d'\^{e}tre for the
presence of an }$H$\textit{-particle in models of selfinteracting massive
vector mesons.}

The application of the SLFT perturbation theory to the Higgs model leads to
other foundational questions whose answer may be trendsetting for the
development of QFT. The basic interaction density $A\cdot AH$ for a $A_{\mu
},H$ particle content (the "ignition" from which the $L,V_{\mu}$ pair
requirement uniquely induces all other contributions) is superrenormalizable
since $d_{sd}(AAH)=d_{cl}=3$. One does not expect that interactions induced by
superrenormalizable couplings lead to higher order counterterms with new
coupling parameters. A $4^{th}$ order confirmation of this expectation does
presently not exist (neither in SLFT nor in the gauge theoretic SSB setting).

Even more important is to find out if $SLFT$ permits an extension to $s\geq2.$
The remarks in section $5.6~$on $s=2$ selfinteractions show that
$d_{sd}(L)=5.$ To conclude that the theory is useless because it violates the
power-counting bound is premature since (previous section) the main reason for
dismissing interaction densities is that they lead to a with perturbative
order increasing number of coupling parameters and not the fact that there are
fields with an increasing short distance scaling degree. For the acceptance of
a model it suffices that its S-matrix is well-defined and that its observables
remain pl with bounded $d_{sd},$ independent of whether the $d_{sd}$ of its sl
interpolating fields increase with perturbative order.

Candidates with $s=2$ potentials $A_{\mu\nu}~$and superrenormalizable
"ignition"$~$of the form $A_{\mu\nu}A^{\mu\nu}H~$or $A^{\mu\nu}A_{\mu}A_{\nu
}~$and induced $L,V_{\mu}$ pairs couplings with $d_{sd}(L)=5~$are expected to
exist. As long as the number of counterterm coupling parameters does not
increase with perturbative order and the physical predictability is maintained
there is no obvious reason for their exclusion of such $L,V_{\mu}$ pairs. Only
further research can resolve these challenging problems. \ 

Can SLFT shed some light on the perplexing question \textit{why GT inspite of
its obvious conceptual shortcomings\footnote{Positivity is an indispensable
property which secures the probability interpretation of quantum theory. }
remained such an amazingly successful theory}? This paradigmatic question may
have a positive answer. The $L^{K},V_{\mu}^{K}~$pair property is a consequence
of the relation $A_{\mu}^{P}=A_{\mu}^{K}+\partial_{\mu}\phi^{P,K}$ where $K$
refers to the Krein space of GT and $\phi^{P,K}$ is a scalar pl "hybrid"
escort which mediates between the $P$ and $K$ formalism \cite{hybrid}.

In the BRST formulation used in \cite{Scharf} the fields act in a
St\"{u}ckelberg- and ghost- extended BRST space. The physical space, to which
the action of $A_{\mu}^{P}$ can be restricted, is defined in terms of BRST
cohomology and observables are defined as objects invariant under the BRST
operation $\mathfrak{s}$ ($\mathfrak{s}O=0$ for observables and $\mathfrak{s}%
S=0$ for the S-matrix).$\mathfrak{~}$

The advantage of the hybrid formulation proposed by Mund \cite{hybrid} is
that, different from the formalism used in \cite{Scharf}, St\"{u}ckelberg- and
ghost- degrees of freedom are not needed. Instead of spaces which are embedded
in the sense of BRST cohomology one deals with factorization through
Gupta-Bleuler subspaces.

An explicit expression for the mixed hybrid escort $\phi^{P,K}$ was recently
calculated by Mund (private communication). The Proca potential lives in the
transverse subspace to the mass shell $p^{\mu}\psi_{\mu}(p)=0$ which is
embedded in the Krein space whereas the living space of fields is the full
4-component Krein space. The triple relation can the be used to define a
$L^{K},V_{\mu}^{K}$ pair which is S-matrix-related to the physical $L^{P}$ formulation.

A favorable situation for studying \textit{infrared phenomena} arises from the
hybrid triple $A_{\mu}^{sl}(x,e)=A_{\mu}^{K}(x)+\partial_{\mu}\phi
^{sl,K}(x,e).~$The $A_{\mu}^{S}~$(without)~lives on the physical subspace of
the Gupta-Bleuler Krein space. All three contributions have a massless limit,
but $\phi^{s,K}$ without the derivative has the typical logarithmic infrared
divergence known from scattering theory of charge-carrying particles. Its
exponential $\exp ig\phi^{S,K}(x,e)$ seems to provide the kind of directional
superselection rule of photon "clouds" whose presence is required by a theorem
\cite{Bu}. This picture is a closer analog of (\ref{scal}) than the $\exp
ig\Phi(x,e,e^{\prime})$ constructed in (\ref{5}).

This hybrid pair description does not only explain the close relation of a
(from ghosts and negative metric St\"{u}ckelberg fields liberated)
Gupta-Bleuler GT with the positivity preserving SLFT, but it also shows that
GT plays a useful constructive role for a better understanding of SLFT in QED.
The hybrid relation reveals that the physical origin of quantum gauge theory
is that one cannot squeeze causally spacetime localized pl vector potentials
into the Wigner momentum space; this is only possible by permitting a
noncompact but still causally separating localization.

In particular it contains informations about the change of the Wigner particle
space for the $\mathcal{B}|_{L}$ operators (previous section) in the massless
limit.$~$Whereas the fields in $B$ live in a Wigner-Fock helicity space, their
interacting images in $\mathcal{B}|_{L}$ act on a larger space for whose
construction one needs to form line integrals on indefinite Gupta-Bleuler
potentials $A_{\mu}^{K}(x)$ (still indefinite) and convert them into complex
exponential fields (the photon clouds) whose associated Hilbert space is
expected to show a similar infrared structure as the exponentials $\exp
ig\varphi(x)$ of the indefinite logarithic divergent massless $d=1+1$ scalar
fields (the $\varphi$-clouds).

The hope is to obtain a spacetime understanding of infrared phenomena
including the large-time behavior which replaces that of the LSZ scattering
theory. This includes the vanishing of scattering between charge-carrying
particles with only a finite number of outgoing photons.

This cannot be described solely in terms of free matter fields, rather the
exponential sl dependent photon cloud fields must play an important role.
Similar to the $\varphi$-clouds in a two dimensional model (\ref{scal}) they
are expected to "soften" the mass-shell singularity and account for the zero
probability for the emission of a finite number of photons in collisions of
charged particles whereas a perturbative expansion which ignores this change
of the mass-shell leads to the logarithmic infrared singularities. As often,
the devil is in the details.

SLFT also suggests that behind confinement there could be a more radical
\textit{off-shell} perturbative logarithmic infrared divergence of massless
selfinteracting gluons. Such off-shell divergences are absent in covariant
gauges of nonabelian GT, but off-shell long distance singular behavior of
self-interacting gluons \textit{in SLFT} may be stronger than in GT
\cite{Bey}\cite{dark}.

Most problems of SLFT remain unsolved; on particular the present state of
knowledge about higher order perturbative renormalization is insuffient. Its
strengths is that the new ideas passed many tests and that the promise to
transcend the conceptual limitations of GT is too tempting to resist.

This may be attributed in part to the fact that its underlying ideas are in
embryo and the number of researchers who know about their existence and
decided to study them is still very small. There is no lack of researchers
working on foundational problems of QFT extending the pioneering work of
Wightman, Haag and others. Most theoreticians use the existing gauge theoretic
formalism to solve problems of high energy particle physics or cosmology.
During the last 5 decades a lot of time has been invested in reseach on
speculative ideas as String Theory, Multiverses, Supersymmetry and alike; the
incentive was obviously to continue the success of the first three decades of
QFT in which such speculative way of proceeding was very successful and which
led to most of our by now household goods.

The lack of any tangible results of these attempts led meawhile to feelings of
somberness. The SLFT raises the question why loose time with speculative ideas
if we still know so little about our most successful theory?

\textbf{Acknowledgements.} This work is part of a joint project whose aim is
to develop a positivity-preserving formulation of causal perturbation theory
QFT which, instead of using indefinite metric and "ghosts", is based on only
physical degrees of freedom. The people presently involved in it are Jos\'{e}
Gracia-Bond\'{\i}a, Jens Mund, Karl-Henning Rehren and Joseph V\'{a}rilly and
their students. For intense and profitable contacts I owe thanks to all of
them. I am also indebted to Michael D\"{u}tsch for helpful comments.

I acknowledge a stimulating correspondence with Edward Witten which started
with his curiosity about the history of Haag duality in connection with a
recent upsurge of interest in the role of causality in problems of
entanglement. SLFT has similar conceptual roots and particle physics would
profit if it could divert some of this new interest in fundamental properties
of QFT. \ Last not least I thank the referee for his constructive proposals.

\end{document}